\documentclass[aps,twocolumn,nofootinbib,showpacs,floatfix]{revtex4}

\usepackage[centertags]{amsmath}
\usepackage{amsfonts}
\usepackage{graphicx}
\usepackage[hypertex]{hyperref}
\usepackage{psfrag}

\newcommand{\bea}{\begin{eqnarray}}
\newcommand{\eea}{\end{eqnarray}}
\newcommand{\be}{\begin{equation}}
\newcommand{\ee}{\end{equation}}

\newcommand{\lan}{\langle}
\newcommand{\ran}{\rangle}
\newcommand{\ti}{\tilde} 

\newcommand{\de}{\delta}

\newcommand{\Ga}{\Gamma}

\newcommand{\Om}{\Omega} 
\newcommand{\si}{\sigma}

\newcommand{\ra}{\rightarrow}

\newcommand{\bean}{\begin{eqnarray*}} 
\newcommand{\eean}{\end{eqnarray*}}

\def\lsim{\raise 0.4ex\hbox{$<$}\kern -0.8em\lower 0.62ex\hbox{$\sim$}} 
\def\gsim{\raise 0.4ex\hbox{$>$}\kern -0.7em\lower 0.62ex\hbox{$\sim$}}

\newcommand{\br}{{\bf r}}

\newcommand{\bq}{{\mathbf q}}
\newcommand{\bse}{\begin{subequations}}
\newcommand{\ese}{\end{subequations}}

\begin{document} 
\title{Quantification of discreteness effects in cosmological N-body simulations: \\ I. Initial Conditions
}
\author{M. Joyce}   
\affiliation{Laboratoire de Physique Nucl\'eaire et de Hautes Energies,  
 Universit\'e Pierre et Marie Curie --- Paris 6, UMR 7585, 
Paris, F-75005, France} 
\author{B. Marcos}   
\affiliation{ ``E. Fermi'' Center, Via Panisperna 89 A, Compendio del 
Viminale, I-00184 Rome, Italy.} 
\begin{abstract}   
\begin{center}    
{\large\bf Abstract}   
\end{center}    
The relation between the results of cosmological N-body simulations,
and the continuum theoretical models they simulate, is currently not 
understood in a way which allows a quantification of N dependent 
effects. In this first of a series of papers on this issue, we consider 
the quantification of such effects in the initial conditions of
such simulations. A general formalism developed in \cite{andrea}
allows us to write down an exact expression for the power spectrum 
of the point distributions generated by the standard algorithm
for generating such initial conditions. Expanded perturbatively in
the amplitude of the input (i.e. theoretical, continuum) power spectrum, 
we  obtain at linear order the input power spectrum, plus two terms which
arise from discreteness and contribute at large wavenumbers.
For cosmological type power spectra, one obtains as expected, 
the input spectrum for wavenumbers $k$ smaller than that characteristic 
of the discreteness. The comparison of real space correlation properties
is more subtle because the discreteness corrections are not 
as strongly localised in real space. For cosmological type  spectra 
the theoretical mass variance in spheres and two point correlation function 
are well approximated  {\it above} a finite distance. For typical initial 
amplitudes this distance is a few times the inter-particle distance, but 
it diverges as this amplitude (or, equivalently, the initial red-shift of 
the cosmological simulation) goes to zero, at fixed particle density. 
We discuss briefly the physical significance of these discreteness 
terms in the initial conditions, in particular with respect to 
the definition of the continuum limit of N-body simulations.
\end{abstract}    
\maketitle   
\date{today}  

\twocolumngrid   
     
\section{Introduction}   
The goal of dissipationless cosmological N-body simulations (NBS) is
to trace the evolution of the clustering of matter under its
self-gravity, starting from a cosmological time at which perturbations
to homogeneity are small at the physically relevant scales (for
reviews, see e.g. \cite{efstathiou_init, couchman,
bertschinger_98}). A fundamental question about such simulations
concerns the degree to which they reproduce the evolution of the
simulated models. The problem arises because the N-body technique,
which simulates a large number of particles evolving under their
self-gravity (with some small scale regularization), is not a direct
discretization of the theoretical models.  The N-body approach is
taken because it is not numerically feasible to simulate, at a useful
level of resolution, the continuum Vlasov-Poisson equations describing
the evolution theoretically.

Since the Vlasov-Poisson system corresponds 
to an appropriately defined N $\rightarrow \infty$ of this particle
dynamics \cite{braun+hepp, spohn}, NBS can be considered as related to the
models in this limit. The problem of discreteness is thus that of
the relation of the results obtained from these simulations, for 
typical statistical quantities characterising clustering, with
those which would be obtained with such a simulation done with 
N $\rightarrow \infty$ particles. 
The existing studies 
of this ``convergence'' problem in the literature  
(e.g. \cite{efstathiou_init, 
splinter, melott_all, discreteness-hamana, diemandetal_2body,
diemandetal_convergence})
are almost exclusively numerical, and consider the stability of
different measured quantities as a function of N. 
In the absence, however, of any analytic understanding of
the possible N dependence of the results, such studies, which 
extend over a very modest range of N, cannot be conclusive.
Different groups of authors have in fact drawn very different 
conclusions about the correctness of results for standard
quantities at smaller scales. Some \cite{splinter, melott_alone} 
even place in question the validity of results for clustering
amplitudes below the initial interparticle distance, while
such results are widely interpreted as physical in almost
all current simulations. Further it is not specified in such 
studies how precisely the limit of large N should be 
taken, i.e., which other parameters 
(e.g. box-size, force softening, initial red-shift)  should 
be kept fixed or varied. These questions are becoming of ever 
greater practical importance as the quantification of the 
precision of results from simulations 
is essential in order to confront cosmological models with a
rich host of observations (see, e.g., \cite{Huterer:2004tr}).

In this paper we address only one simple aspect of this problem:
the relation between the discretized initial conditions (IC)
of an NBS and the IC of the corresponding theoretical model. 
More specifically we study and quantify analytically the 
differences between the two for the two point correlation properties,
in real space and reciprocal space, in the infinite volume limit 
(at a fixed particle density). The discreteness effects, i.e.,
the differences between the continuous theoretical IC and 
the discrete IC of the actual NBS are then terms which depend
on the particle density. We study these terms and their 
relative importance for different theoretical IC.
We also study and check our analytic results with the
aid of numerical simulations, performed in one dimension because
of the modest numerical cost of calculating the ensemble average
using a large number of realisations. We underline that our analytic 
results apply to the infinite volume
limit, i.e., to the case that the size of the simulation box
goes to infinity at fixed particle density. They thus do not
include effects associated with the finite size of the box.
Such effects, which are quite distinct from those studied
here, have been studied extensively elsewhere
(see e.g. 
\cite{gelb-bertschinger, colombietal1, colombietal2, colombietal3, pen,
sirko}), both analytically
and numerically.

The direct motivation for this work on IC in NBS came originally 
from some {\it numerical} studies of these issues by two groups
\cite{Baertschiger:2001eu, alvaro_knebe, Baertschiger:2003jt,
alvaro_knebe2}, who have drawn  quite different conclusions
about the accuracy with which the IC produced by the canonically 
used algorithm in NBS represent the input IC \footnote{A more detailed
account is given in the conclusions section below.}.
The results in this paper, which are essentially analytic,
clarify the issues underlying the discussion in and differences 
between these numerical studies. Our conclusions
are consistent with the findings of both sets of authors,
and explain the differences.
In short the authors of \cite{alvaro_knebe, alvaro_knebe2} are
correct that certain real space properties, notably the  mass
variance in spheres, are in fact reasonably well 
represented for typical IC in NBS. The authors of
\cite{Baertschiger:2001eu}, however, are correct in
diagnosing the important systematic differences between the
actual and theoretical correlation properties in
real space.  Indeed one of our main findings is that
there is a very non-trivial difference between the two spaces:
while the discreteness of the underlying particle distribution
is strictly localized in reciprocal space, this is not the case
in real space. The result is that, in the  limit of
low amplitude initial density fluctuations --- or, equivalently,
high initial redshift for the simulation --- the correlation
properties of the input theoretical model are approximated
well only in reciprocal space. Taking instead the limit that 
the particle density goes to infinity at fixed amplitude, the 
theoretical correlation properties are recovered in both
real and reciprocal space, provided an appropriate cut-off
is imposed at large wavenumber in the input PS.

The implications of these results for what concerns the 
agreement between an {\it evolved} NBS and the
evolved theoretical model of which it is the discretization 
is beyond the scope of this paper.
In a forthcoming paper \cite{discreteness2_mjbm}, treating
discreteness effects up to shell crossing, we will see that
the evolution of an NBS deviates arbitrarily from its continuum 
counterpart as the initial red-shift increases at fixed particle 
density, while keeping the amplitude fixed at increasing 
particle one approximates increasingly well the continuum
(fluid limit) evolution. Thus the results we find for the
initial conditions here do indeed turn out to have physical 
significance for the question of discreteness effects in
the evolved simulations.

The algorithm used to generate IC in NBS which we analyze 
is in fact well defined, in the infinite volume limit, only
for a certain range of asymptotic behaviors of the input 
theoretical PS. We specify here carefully these limits. 
While it turns out that they are not particularly relevant 
to current cosmological models, they are of interest in other
contexts in which this algorithm may be used, notably in the
study of gravitational clustering from other classes of
initial conditions (e.g. \cite{melott+shandarin_k4,
bagla+padmanabhan_97}). These properties are 
also of interest 
in the context of statistical physics, where the problem
of ``realizability'' of point processes is studied (see e.g.
\cite{costin+lebowitz, crawford+torquato+stillinger, 
ucheetal_2006}). Specifically we find that the algorithm 
has interesting limits for the case of very ``blue'' input PS: 
for the case of spectra with a small $k$ behavior 
proportional to $k^{n}$, and $n>1$, the real space
variance is never that of the input model, while 
for $n>4$ the reciprocal space representation is 
never faithful either.  
 
The paper is organized as follows. 
In the section which follows we briefly review the standard method
for setting up IC for cosmological simulations, using the Zeldovich
approximation. This also sets conventions for notation in the rest of the paper.
In Sect. \ref{IC_exact_kspace}  we analyze the PS of
the configurations of points generated in this way,
comparing it with the PS of the input theoretical
model. To obtain these results we use a very general exact result 
derived in \cite{andrea},
which gives the two point properties of a point distribution generated
by superimposing an arbitrary correlated displacement field on an
arbitrary initial stochastic point distribution. 
In the following section we consider how these 
properties described in $k$-space translate into  
real space. Specifically we present 
a general qualitative analysis of the 
relation between the two 
point correlation properties of the IC
and those of the input models. We treat
specifically  the mass variance in spheres,
and the reduced two point correlation function.
For the latter case the comparison of the theoretical
model and full IC is more difficult, because 
of the complicated non-monotonic form of the 
correlation function of the underlying point
distributions. In the following section we illustrate,
and verify our results quantitatively, using
one dimensional numerical simulations. We choose
to work in one dimension for numerical economy,
and because all the pertinent questions can be
posed equally well and answered in this case\footnote{The
same is evidently not true when considering
discreteness effect in the dynamics.}. 
In the final section we summarize our results,
discuss what conclusions can be drawn concerning
the papers mentioned above which motivated the 
present study, and finally briefly comment on
the physical significance of our results, which
will be further developed in the companion
paper \cite{discreteness2_mjbm}. Several technical
details in the paper, notably concerning the 
perturbative expansion of the exact expression
for the PS of the generated IC, are given in
three appendices.

\section{Generation of IC using the Zeldovich approximation} 
\label{IC-ZA}   

The method which is used canonically for the generation of IC in 
cosmological NBS is based on the so-called Zeldovich approximation
(ZA)\cite{zeldovich_70}. 
It may be derived at linear order in a perturbative treatment
of the equations describing a self-gravitating fluid 
in the Lagrangian formulation\cite{buchert2}.
It relates the initial position $\mathbf q$
of a fluid element to its final position\footnote{We do not make the 
distinction here between physical and
comoving coordinates, and do not write the associated time 
dependent factors. Since we will analyze only IC
for density fluctuations (and not velocities) these are
not relevant details, and so we omit them for simplicity.} $\mathbf{x}$ through the
expression 
\begin{equation}
\label{displa}
\mathbf{x}(\mathbf q,t)=\mathbf{q}+f(t)\mathbf u(\mathbf q)\,,
\end{equation}
i.e., it expresses the displacement of a particle as a 
separable function of the initial position $\mathbf{q}$ and 
the time $t$. The function $f(t)$ is equal, up to an
arbitrary normalization, to the growth factor of 
density perturbations derived in linear perturbation 
theory (see below). The  vector field 
$\mathbf{u}(\mathbf{q})$ is thus proportional to both the velocity
and acceleration of the fluid element, and with a suitable
normalization it can thus be taken to satisfy
\begin{equation}
\label{velocity-ZA}
\mathbf u(\mathbf q)=-\mathbf\nabla\cdot\Phi(\mathbf q)
\end{equation}
where $\Phi(\mathbf q)$ is the gravitational potential at the initial
time created by the density fluctuations. To set up IC representing
a density field, one thus simply determines the associated potential 
through the Poisson equation and infers the appropriate displacements
[and velocities, given $f(t)$] using Eq.~(\ref{displa}) to apply to
a set of points representing the unperturbed fluid elements.

We note that to understand this algorithm for generating
IC representing a continuous density field it is not in fact necessary
to invoke the ZA, nor anything which has specifically to do with
gravity. These latter are relevant only for the determination of
the velocity field. The only relation needed is in fact the
continuity equation, which relates the velocity (and thus
displacement) field in a fluid to the density fluctuations. 
At leading order in the density fluctuations it gives
\be
\label{delta_rho}
\delta\rho(\mathbf x)=-\mathbf\nabla\cdot\mathbf u(\mathbf x).
\ee
where the density fluctuation $\delta\rho(\mathbf x)$
is defined by
\be
\delta\rho(\mathbf{x})=\frac{\rho({\mathbf x})-\rho_0}{\rho_0},
\ee
$\rho(\mathbf{x})$ is the (continuous) density field,
$\rho_0$ the average density, and $\mathbf u(\mathbf x)$ is
the displacement field. By inversion one can determine
a displacement field which gives a desired density field.
If one {\it assumes} further that the former is curl-free,
and thus derivable as the gradient of a scalar field,
one obtains a unique prescription for the displacement
field which is identical to that given by the ZA as
described above.

In cosmological models the starting point for IC is not
a specific density field, but a power spectrum (PS) of
density fluctuations. The latter is defined as 
\be
P(\mathbf k)=\lim_{V\to\infty}\frac{\lan {|\de\hat\rho(\mathbf k)|^2}\ran}{V}\,,
\ee
where $\langle\dots\rangle$ denotes the average over an ensemble 
of realizations and $\hat\de\hat\rho(\mathbf k)$ denotes the Fourier 
transform (FT) of $\rho(\mathbf x)$ defined as
\be
\label{FT-def}
\de\hat\rho(\mathbf k)=\int_V d^dx \de\rho(\mathbf{x}) e^{-i\mathbf{k}\cdot\mathbf{x}}.
\ee
It follows then from Eq.~(\ref{delta_rho}) 
that
\be
\label{PS_u}
P(\mathbf k)= k_i k_j \hat{g}_{ij} (\mathbf{k})
\ee
where 
\be
\label{g(k)def}
g_{ij} (\mathbf{k}) = \lim_{V \to \infty}
\frac{\langle\hat{u}_i (\mathbf{k}) \hat{u}_j^* (\mathbf{k}) \rangle}{V}
\ee
and  $\mathbf{\hat u}(\mathbf k)$ is the Fourier transform (FT) 
of the vector field $\mathbf{u}(\mathbf{q})$.
Assuming that the latter is derived from a scalar potential
as in Eq.~(\ref{velocity-ZA})  we have 
\be
\label{g(k)form}
\hat{g}_{ij} (\mathbf{k}) =\hat{k}_i \hat{k}_j \hat{g} (k)
\ee
where $\hat{g} (\mathbf{k})=Tr[\hat{g}_{ij} (\mathbf{k})]$ 
is a function of $k=|\mathbf{k}|$ only because  
the stochastic process is assumed to be 
statistically homogeneous and isotropic, and
$\mathbf{\hat k}=\mathbf{k}/|\mathbf{k}|$. We thus have 
\be
P(\mathbf k)=  k^2 \hat{g} (k)=  k^4 P_\Phi (k) 
\label{densityPS-displacementsPS}
\ee
where $P_\Phi (k)$ is the PS of the fluctuations
in the scalar field, i.e.,
\be
\label{PS-phi}
P_\Phi (k)= \lim_{V \to \infty}
\frac{\langle|\Phi(\mathbf{k})|^2 \rangle}{V}.
\ee

If one considers now a displacement field which varies
as a function of time as in Eq.~(\ref{displa}), it follows 
that the PS of density fluctuations is proportional to the 
square of the function $f(t)$. For a self-gravitating
fluid such a behavior applies, and thus one can
determine the function $f(t)$  for the determination
of the velocities\footnote{Normally $f(t)$ is chosen so
that density perturbations are in the pure growing mode
in which the velocity field is parallel to the displacement
field.}. Indeed Zeldovich originally proposed
his approximation as an ansatz, on the basis that
Eq.~(\ref{displa}) implies the correct evolution of
the density fluctuation in linearized Eulerian
theory. The power of the ZA is that it can be applied
well beyond the regime of validity of Eulerian
perturbation theory, to which it matches at
early times.

To set up IC for the $N$ particles of a cosmological 
NBS the procedure is then \cite{white_leshouches, efstathiou_init}:
\begin{itemize}
\item to set-up a ``pre-initial'' configuration of the
$N$ particles. This configuration should represent
the fluid of uniform density $\rho_0$. The usual choice
is a simple lattice, but a commonly used alternative
\cite{white_leshouches} is an initial ``glassy'' configuration 
obtained by evolving the system with negative gravity (i.e. a 
Coulomb force) with an appropriate damping.
\item given an input theoretical PS $P_{th}(k)$, 
the corresponding displacement field in the ZA
is applied to the ``pre-initial'' point distribution.
The cosmological IC are usually taken to be Gaussian, and 
the displacements are determined by generating 
a realization of the gravitational potential 
\begin{equation}
\label{potential}
\Phi(\mathbf q)=\sum_{\mathbf k} a_{\mathbf k}\cos(\mathbf k \cdot\mathbf q)+b_{\mathbf
k}\sin(\mathbf k \cdot \mathbf q)
\end{equation}
with
\be
\label{coeff}
a_{\mathbf k}=R_1\frac{\sqrt{P_{th} (k)}}{k^2},\quad b_{\mathbf
k}=R_2\frac{\sqrt{P_{th}(k)}}{k^2}\,, 
\ee
where $R_1$ and $R_2$ are Gaussian random numbers of
mean zero and dispersion unity. 
From Eq.~(\ref{densityPS-displacementsPS}) we
see that this corresponds to generating a 
realization of a stochastic displacement field 
with PS 
$\hat{g}_{ij} (\mathbf{k})$ as in Eq.~(\ref{g(k)form})
and 
\be
\label{PS-disp} 
\hat{g}(k)= P_{th}(k)/k^2 \,.
\ee

\end{itemize}

\section{
Analytic results in $k$-space}
\label{IC_exact_kspace}   

The configuration (or ensemble of configurations) generated
by the method outlined in the previous section has PS
given through Eq.~(\ref{densityPS-displacementsPS}), 
and thus equal to the theoretical PS $P_{th}(k)$, up to
the following approximations:
\begin{itemize}
\item The system is considered as a continuous fluid.
Thus we expect the exact PS of the 
(discrete) particle distribution to differ by terms
which come from the ``granularity'' (i.e. particle-like)
nature of this distribution.

\item The calculations are performed at leading order
in the amplitude of the density fluctuations, or
equivalently, at leading order in the gradient of the 
displacements (cf. Eq.~(\ref{delta_rho})). We would 
thus anticipate that the exact PS of the generated 
configurations will have
corrections which are significant for $k$ larger than
the inverse of a scale characterising the 
amplitude of the input PS. 
\end{itemize}

The rest of the paper is principally focussed on 
the consideration of the differences arising from the
first point between the theoretical PS $P_{th}(k)$ and 
the exact PS (which we will simply
denote $P(\mathbf{k})$) of the distribution generated
by the algorithm described in the previous 
section\footnote{Note that the full PS is assumed
to be a function of $\mathbf{k}$, as it will not in general
share the statistical isotropy and homogeneity of
the theoretical PS (which makes it a function only
of $k=|\mathbf{k}|$).}. We refer the reader to
\cite{scoccimarro_transients_98, valageas_IC_03}
for analyses of the second point, i.e. of corrections
coming from the use of the leading order ZA. These latter
studies work in the continuum limit, and so completely decouple 
the problem of non-linear corrections from the effects of
discreteness studied here. 

\subsection{General results}
\label{subsection-general-results}
The starting point for our analysis is a result derived
in \cite{andrea}. One considers, in $d$ dimensions, 
the application of a displacement field 
$\mathbf{u}(\mathbf{r})$ to a generic point distribution. The
latter is taken to have PS $P_{in} (\mathbf{k})$ and
correlation function $\ti\xi_{in} (\mathbf{r})$, given
by the inverse Fourier transform 
\begin{equation}
\label{definition_xi1}
\ti\xi_{in}(\mathbf r)=\frac{1}{(2\pi)^d}\int d^dk\,e^{-i\mathbf k \cdot \mathbf r}P_{in}(\mathbf k),
\end{equation}
where the integral is over all space. The displacement 
field $\mathbf{u}(\mathbf{r})$ 
is assumed to be a realization of a {\it continuous} stochastic process,
which is statistically homogeneous. 
An exact calculation  \cite{andrea} gives
that the PS of the distribution obtained in this way
may be written as
\be 
\label{PS_andrea}
P(\mathbf k)=\int d^dr e^{-i\mathbf k \cdot \mathbf r}\hat{p}(\mathbf k;\mathbf
r)\left(1+\tilde\xi_{in}(\mathbf r)\right)-(2\pi)^d\delta(\mathbf k). 
\ee
where 
\begin{equation}
\label{hatp}
\hat{p}(\mathbf k;\mathbf r)=\int d^ds\,e^{-i\mathbf k \cdot \mathbf s}p(\mathbf s;\mathbf r),
\end{equation}
and $p(\mathbf s;\mathbf r)$ is the probability that 
two particles with a separation $\mathbf r$ undergo a relative 
displacement $\mathbf s$. 

We note that our choice of notation here follows also
that of \cite{sb95} (rather than that of \cite{andrea}).
In this work (see also \cite{taylor-hamilton}) an expression for the PS 
generated by displacements given by the ZA is derived,
for the case of a continuous fluid. The general expression
given is exactly that obtained by setting $\ti\xi_{in}(\mathbf r)=0$
in (\ref{PS_andrea}). This extra term in our expression
arises because we do not make the approximation of
treating the ``pre-initial'' configuration
as a continuous uniform background.
We note that this additional term contains not just the 
effect of taking into account the correlations in 
the ``pre-initial'' configuration, but also includes
more generally all the effects of the discreteness 
of the (``pre-initial'' and final) distribution. 
In this respect we note that the correlation 
function $\ti\xi_{in}(\mathbf r)$ for the ``pre-initial'' 
distribution contains generically a delta-function
at $r=0$, which is characteristic of its discreteness,
as well as a non-singular function which describes
correlations (for a detailed discussion see \cite{book,glasslike}). 

From the definition of $p(\mathbf{s},\mathbf{r})$ it follows 
that 
\be
\label{p(k,r)}
\hat{p}(\mathbf{k},\mathbf{r}) = 
\int {\cal D}\mathbf{u} \, {\cal P}(\{\mathbf{u}({\mathbf{r}})\})  
\,e^{-i\mathbf{k}\cdot [\mathbf{u}(\mathbf{r})-\mathbf{u}(\mathbf{0})]}
\ee
where the functional integral is over all possible displacement
fields weighted by their probability ${\cal P}(\{\mathbf{u}(\mathbf{r})\})$. For
a displacement field which is (i) Gaussian, and (ii) statistically
isotropic (as well as homogeneous) it is then simple to show 
\cite{andrea} that 
\begin{equation}
\label{p_g}
\hat{p}(\mathbf{k},\mathbf{r})= e^{-k_i k_j d_{ij}(\mathbf{r})}
\end{equation}
where a sum is implied over the labels $i$ and $j$, and 
\be
\label{definition-d}
d_{ij}(\mathbf{r})\equiv g_{ij}(0) - g_{ij}(\mathbf{r})\,,
\ee
where 
\be
\label{disp-corrfn}
g_{ij}(\mathbf{r}) = \lan u_i(0) u_j(\mathbf{r}) \ran
\ee
is the (matrix) two-displacement correlation function.
We note that the scalar function 
\be
\label{disp-corrfn-trace}
g(r) = Tr ( g_{ij}(\mathbf{r})) = \lan \mathbf{u} (0) \cdot \mathbf{u}(\mathbf{r}) \ran
\ee
is simply the inverse FT of $\hat{g}(k)$ defined above (and, by statistical 
isotropy, a function only of $r=|\mathbf{r}|$), and that 
$g(0)=\langle u^2\rangle$ is the variance of the displacement field. 
We have that
\be
d_{ij}(\mathbf{r})= \frac{1}{2} \langle \left[u_i(0) - u_i({\mathbf r})\right]
\left[ u_j(0) - u_j({\mathbf r}) \right] \rangle \,,
\label{definition_d2}
\ee
i.e., it is proportional to the correlation matrix of the 
{\it relative} displacements.

Substituting (\ref{p_g}) in (\ref{PS_andrea}), we obtain
\begin{eqnarray}
\label{P_3d}
\nonumber P(\mathbf{k})&=& \int_\Om d^dr
e^{-i\mathbf k\mathbf{r}}
e^{-k_i k_j d_{ij}(\mathbf{r})}
\left(1+\tilde\xi_{in}(\mathbf{r})\right)\\ &-&(2\pi)^d\delta(\mathbf k).
\end{eqnarray}
It will be useful for our discussion to break this
expression into two pieces,  
$P(\mathbf{k})=P_{\rm c} (\mathbf{k})+P_{\rm d} (\mathbf{k})$,
written in the form 
%
\begin{subequations}
\begin{align}
\label{P_cont}
P_{\rm c}(\mathbf{k})&= \int_\Om d^dr
e^{-i\mathbf k\mathbf{r}}
\left(e^{-k_i k_jd_{ij}(\mathbf{r})}-1\right)\\
\label{P_disc}
\nonumber
P_{\rm d}(\mathbf{k})&= P_{\rm in}(\mathbf{k})
+ \int_\Om d^dr
e^{-i\mathbf k\mathbf{r}}
\left(e^{-k_i k_j d_{ij}(\mathbf{r})}-1\right)
\tilde\xi_{in}(\mathbf{r}) \,.\\
\end{align}
\end{subequations}
The first term Eq.~(\ref{P_cont}) is the ``continuous'' piece
of the generated PS (identical, as discussed above to
that given in \cite{sb95}), and the second term  Eq.~(\ref{P_disc})
is the contribution coming from the discreteness. 

\subsection{Application to cosmological IC}
\label{subsection-Application to cosmological IC}

In the algorithm used to generate cosmological NBS, we have
seen that the FT of $g_{ij} ({\mathbf r})$ is 
[cf. Eqs.~(\ref{g(k)form}) and (\ref{PS-disp})] given by
\begin{equation}
\label{cosmoIC_g}
\hat{g}_{ij} (\mathbf{k}) =\hat{k}_i \hat{k}_j \frac{P_{\rm th} (k)}{k^2}\,.
\end{equation}
Expanding the exponential factor in Eq.~(\ref{P_cont}) and 
(\ref{P_disc}) in power series,
we can thus obtain expressions for $P_{\rm c}(\mathbf{k})$ and 
$P_{\rm d}(\mathbf{k})$ at each order 
in powers of $P_{\rm th} (k)$.
At zero order we have evidently
\begin{equation}
P^{(0)}_{\rm c}(\mathbf{k})=0 \qquad P^{(0)}_{\rm d}(\mathbf{k})=P_{\rm in}(\mathbf{k})
\label{PS_order0_cont+disc} 
\end{equation}
and, at linear order,
\begin{subequations}
\begin{align}
\label{PS_order1_cont} 
P^{(1)}_{\rm c}(k)=P_{\rm th} (k)& \\
\nonumber
P^{(1)}_{\rm d}(k)=\frac{k^2}{(2\pi)^d}\int_\Om d^dq& 
(\mathbf{\hat k}\cdot \mathbf{\hat q})^2
\frac{P_{th} (q)}{q^2} \\
&\times[P_{in}(\mathbf k+\mathbf q)
-P_{in}(\mathbf{k})]
\label{PS_order1_disc} 
\end{align}
\end{subequations}
To this order the PS of the generated distribution is thus
the sum of the input theoretical PS and two discreteness
terms:  the PS of the ``pre-initial'' (i.e. lattice or glass) 
distribution and a second term which is a convolution of the 
input PS and the ``pre-initial'' PS. At next order in the
expansion (i.e. at second order in 
$P_{\rm th}(k)$) we will obtain both further discreteness
corrections, and corrections which survive in the limit
in which we neglect discreteness completely. This result
is in line with what we anticipated at the
beginning of this section. 

\subsection{Domain of validity of the expansion}
\label{subsection-Domain of validity of the expansion}

We have implicitly assumed above that the expansion 
we have performed is well defined \footnote{We note 
that we have also assumed Gaussianity 
in deriving Eq.~(\ref{P_3d}). This is not in fact a necessary 
condition to obtain Eqs.~(\ref{PS_order0_cont+disc}-\ref{PS_order1_disc}).
Making instead only the assumption that $d_{ij}({\mathbf r})$ is bounded,
it is easy (see also \cite{andrea}) to recover
the same result  directly from an expansion of Eq.~(\ref{p(k,r)}).}.
This assumption corresponds to that of finiteness
of various integrals of the input PS 
$P_{\rm th}(k)$. If the latter function is well-behaved, 
this corresponds to constraints on its 
asymptotic properties,
at small and large $k$. To determine these constraints 
we consider a PS of the form
\be
\label{PS-generalIC}
P_{th}(k) = Ak^n f(k/k_c)
\ee
where $A$ and $n$ are constants, and $f(x)$ is a 
function which interpolates between unity 
for $x \ll 1$ and zero for $x \gg 1$, i.e., which 
may act as a cut-off for $k > k_c$. In the use of
this algorithm in cosmological simulations, for reasons
which we will discuss further below, a very
abrupt (usually top-hat) such cut-off is always 
imposed at wavenumbers of order the inverse of the 
scale characteristic of the interparticle separation
\footnote{In this case the cut-off imposed in 
simulations, as explained below, is 
actually a function of {\bf k }.}. 
Thus we will consider only the constraints at small 
$k$, i.e., the lower bound placed on the index $n$. 

Firstly we note that, using Eqs.~(\ref{definition-d}) 
and ~(\ref{cosmoIC_g}), it is simple to show 
that $d_{ij}(\br)$ is well defined only if 
\be
\label{PS-convergence-finite-var-smallk}
\lim_{k \ra 0} k^{d} P_{th}(k) =0\,,  
\ee
i.e., if $n > -d$ in Eq.~(\ref{PS-generalIC}).
This is a condition which is always satisfied
in cosmological models, as it follows from
the finiteness of the one-point variance of 
the theoretical density fluctuations
\footnote{The one point variance of 
density fluctuations is equal to
$\ti\xi_{th}(0)$, which is proportional
[cf. Eq.~(\ref{definition_xi1})] to the 
integral of $P_{th}(k)$.}.

In App.~\ref{expansion-Pc} 
we analyse in detail the full expansion of $P_c(k)$ 
to all orders in $A$, separating  two different cases:
(i) $-d< n <-d+2$, in which the one point variance
$g(0)$ is infinite, and (ii) $ n>-d+2$, in which
case $g(0)$ is finite. From the expansions in
each case one can infer the following: 
\begin{itemize}
\item For $-d<n<-d+2$ 
the leading non-zero term, equal to $P_{th}(k)$,
approximates well the full $P_{\rm c}(\mathbf{k})$ for 
the range of $k$
in which\footnote{In the cosmological literature
$\Delta^2(k)$ is canonically defined with a numerical
prefactor so that $\ti\xi (0)=\langle \delta \rho^2 (0) \rangle
=\int \Delta^2 (k) d(\ln k)$. Given that the resultant
factor depends on the dimension $d$ we will not include it here.}
\be
\Delta_{\rm th}^2(k) \equiv k^d P_{th} (k) \ll 1\,.
\label{validity-criterion-2}
\ee

\item 
For $ -d+2 < n <4$ the criterion to satisfy the same
condition is:
\be
\label{validity-criterion-1}
k^2 g(0) = k^2 \langle u^2 \rangle \ll 1\,.
\ee

\item
For $n \geq (-d+2)+d/2$ there is, at next order in the
expansion of $P_{\rm c}(k)$, a correction proportional 
to $k^4$. This implies that, for $n \geq 4$ the leading
term $P_{th}(k)$ is never well approximated  
at asymptotically small $k$ by the PS of the generated IC.

\end{itemize}

To analyse the expansion of the discreteness
contribution $P_{\rm d} (\mathbf{k})$ we need to 
specify the ``pre-initial'' distribution. 
It is evident however that generically it is at least
as convergent as than that of $P_c(k)$, since
Eq.~(\ref{P_disc}) contains in the integrand 
simply an extra factor of $\tilde\xi_{in}(\mathbf{r})$, which is
typically smaller than unity and decreasing
at large separations.
For a Poisson distribution of number density $n_0$,
for example,  one has $\tilde\xi_{in}(\mathbf{r})=\frac{1}{n_0} \delta (\mathbf{r})$
(where $\delta(\mathbf{r})$ is a Dirac delta function in
$d$-dimensions), and therefore the expansion becomes
trivial with $P_{\rm d}(k)=P_{\rm in}(k)=\frac{1}{n_0}$  
at all orders\footnote{At leading order in the amplitude
of the input theoretical PS  $P_{\rm th}$ one therefore has
$P(k) = \frac{1}{n_0} + P_{th} (k)$.
Thus for an exponent $n < 0$ in
(\ref{PS-generalIC}) one will
have $P(k) \approx P_{th} (k)$ for all
$k \ll (An_0)^{1/n}$. 
For $n>0$,
on the other hand, one can have 
$P(k) \approx P_{th} (k)$ at most in 
an intermediate range of $k$: at small
$k$ the Poisson variance of the ``pre-initial''
distribution will always dominate.}.
In cosmological NBS the ``pre-initial'' distribution, as 
we have discussed, is usually taken to be a simple
lattice or glass. We will see below that for the
case of the lattice the coefficients of the expansion
are sums which are regulated at small $k$, by
the Nyquist frequency of the lattice (defined below).
For the case of the glass, or indeed any distribution 
with an analytic $P_{\rm in}(\mathbf{k})$, we limit 
ourselves to an analysis of the integral coefficient 
of the leading term in Eq.~(\ref{PS_order1_disc}).
It is simple to see, by Taylor expanding  the
expression inside the square brackets at small $q$, that the
finiteness requires only the integrability of
$P_{\rm th}(q)$ at small $q$. This coincides
precisely with the condition 
Eq.~(\ref{PS-convergence-finite-var-smallk}).
We expect that  
$P_{\rm d} (\mathbf{k}) \approx P^{(0)}_{\rm d} (\mathbf{k})+P^{(1)}_{\rm d}
(\mathbf{k})$ will thus also be satisfied when 
Eq.~(\ref{validity-criterion-2}) applies.
We will verify below with numerical simulations that this
is indeed the case.

We note that the condition Eq.~(\ref{validity-criterion-2}) for the 
validity of the perturbative expansion at a given $k$ 
is one which could be guessed from the simple continuum derivation 
using Eq.~(\ref{delta_rho}), in which the expansion
parameter is the amplitude of the theoretical density
fluctuation: $\Delta_{\rm th}^2 (k)$ is 
just a dimensionless measure of the amplitude of the
density fluctuations in the theoretical model arising 
from wavenumbers around $k$~\footnote{Consistent with 
with Eq.~(\ref{delta_rho}), this condition
for the validity of the expansion can be stated 
equivalently in terms of the boundedness of
the dimensionless quantity 
$\frac{|\langle\left[u_i(0) - u_i({\mathbf r})\right]
\left[ u_j(0) - u_j({\mathbf r}) \right] \rangle|}{r^2} 
\label{validity-condition-2}$,
i.e., of the ``gradient'' of the displacement 
fields. We note that in a first version of the 
paper, a stronger condition was given for the validity
of the expansion, $n > -d+2$. This corresponds to 
the condition that variance of the displacement field
be finite. While this stronger condition is assumed 
in the derivations in \cite{andrea}, and 
notably in arriving at Eq.~(\ref{p_g}), it is not a 
necessary condition for the validity of the method.
We thank an anonymous referee for pointing out this
error.}. Further we show in App.~\ref{expansion-Pc} 
that if condition Eq.~(\ref{validity-criterion-1})
is fulfilled for any $k<k_c$,  then Eq.~(\ref{validity-criterion-2})
is also. The two conditions are in fact essentially
equivalent in the case that a cut-off is imposed 
as typically is done the cosmology.


\subsection{The leading non-trivial discreteness correction}
\label{subsection-The leading non-trivial discreteness correction}

Let us now analyse in more detail the leading contribution 
to the generated PS arising from discreteness, i.e., the expression
which we have denoted above by $P_{\rm d}^{(1)}(\mathbf{k})$~\footnote{In
Appendix~\ref{Discreteness corrections to the PS} we present some
further analysis of the full expansion of Eq.~(\ref{P_cont}), providing
analytical expressions for some specific cases.}. We note 
[cf. Eqs.~(\ref{PS_order1_cont}-\ref{PS_order1_disc})] that this 
term arises at the same order as the input PS in the
perturbative expansion, i.e., at linear order in the amplitude
of the input theoretical PS. We consider the specific case of 
a ``pre-initial'' distribution which is a simple cubic lattice. 
Its PS is   
\be
\label{PS_lattice1}
P_{in}(\mathbf k)=(2\pi)^d \sum_{\mathbf h\ne0}\delta(\mathbf k-\mathbf h)
\ee
where the sum over $\mathbf h$ is over all the vectors of the
reciprocal lattice, i.e., $\mathbf h=\mathbf{m} (2\pi/\ell)$, where
$\ell$ is the lattice spacing and $\mathbf{m}$ is a vector of
non-zero integers. The minimal value of $|\mathbf h|=2\pi/\ell$,
is the {\em sampling frequency} $k_s$ of the lattice, equal
to twice the {\em Nyquist frequency}, which we will 
denote $k_N$ (and $k_N=\pi/\ell$). It is instructive to
rewrite the first order term
Eq.~(\ref{PS_order1_disc}) in the form

\begin{figure}
\resizebox{8cm}{!}{\includegraphics*{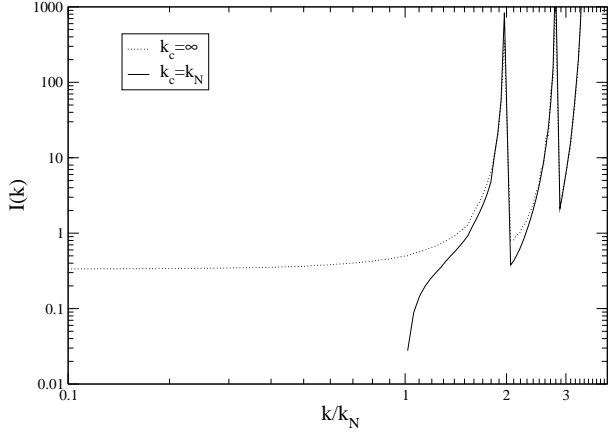}}
\caption{Integral $I(\mathbf{k})$ of Eq.~\eqref{PS_order1_disc_simplified_sum},
in three dimensions and averaged over shells of $|\mathbf{k}|$, for an input 
PS $P_{th}(k)$ as in Eq.~(\ref{PS-generalIC}) with $n=-2$ and 
(i) $f(k)=1$, i.e., without a cut-off (dotted line), and 
(ii)  $f(k)=\Theta(k_N-k)$, i.e., an abrupt (Heaviside
step function) cut-off at $k=k_N$ (solid line). In the former case 
we see that $I(\mathbf{k})$ is approximately constant for $k < k_N$, and
therefore the leading discreteness term $P_d^{(1)}(k) \sim k^2$
for $k < k_N$. In the latter case, the discreteness term contributes
only for $k>k_N$. Such a cut-off is usually employed in the use 
of this algorithm in cosmological simulations.
\label{iofk1}}
\end{figure}

\be
\label{PS_order1_disc_simplified}
P_{\rm d}^{(1)}(k)= Ak^2k_N^{n-2} I(\mathbf{k}) 
\ee
where
\begin{eqnarray}
I(\mathbf{k})\equiv \frac{1}{(2\pi)^d}\int {d^dq} (\mathbf{\hat k} \cdot \mathbf{\hat q})^2
\left(\frac{q}{k_N}\right)^{n-2} f(q/k_c) \nonumber \\
\times [P_{in}(\mathbf k+\mathbf{q})-P_{in}(\mathbf k)]
\label{PS_order1_disc_simplified}
\end{eqnarray}
is dimensionless. Since $P_{in}(\mathbf{k})=0$
for $\mathbf{k}\neq \mathbf{h}$, we therefore have, 
at linear order in our expansion in powers of 
the input PS, that, 
\begin{eqnarray}
P(\mathbf{k})&=&P_{\rm th}(k) + P^{(1)}_d (\mathbf{k})
\nonumber \\ 
&=&Ak^n f(k/k_c) + A k^2k_N^{n-2} I(\mathbf{k})  
\label{PS_smallk-lattice}
\end{eqnarray}
where 
\be
I(\mathbf{k})= 
\sum_{\mathbf h\ne\mathbf 0} \frac{[\mathbf{\hat k}\cdot (\mathbf{h}-\mathbf{k})]^2}{|\mathbf{h}-\mathbf{k}|^2}
\left(\frac{|\mathbf{h}-\mathbf{k}|}{k_N}\right)^{n-2} 
f(\frac{|\mathbf{h}-\mathbf{k}|}{k_c}) 
\label{PS_order1_disc_simplified_sum}
\ee
for $\mathbf{k} \neq \mathbf{h}$.

\begin{figure}
\resizebox{8cm}{!}{\includegraphics*{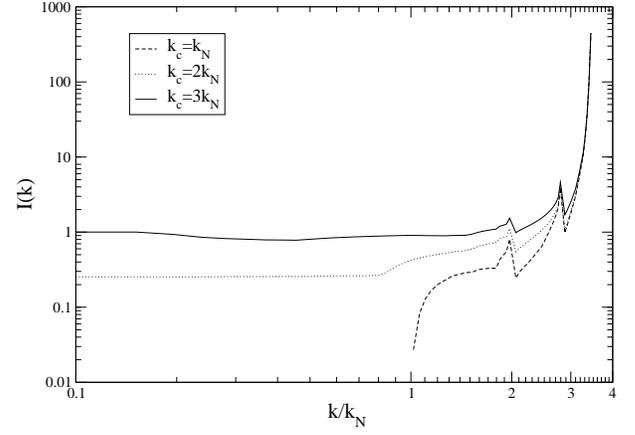}}
\caption{As in previous figure, but now for $n=0$ and 
a top-hat cut-off function, for three different values of 
the cut-off. We see that as the cut-off increases the 
amplitude of $I(k)$ does so (corresponding to the
UV divergence of the sum Eq.~(\ref{PS_order1_disc_simplified_sum})
for $n>-1$ in three dimensions). We see again that for a cut-off at 
$k_N$ the  leading order discreteness term contributes only for $k>k_N$,
while for larger cut-off we have aliasing effects
which manifest themselves in the appearance of the term
$P_d^{(1)}(k) \sim k^2$ for $k < k_N$.
\label{iofk2}}
\end{figure}

The second (discreteness) term in Eq.~(\ref{PS_smallk-lattice})
includes explicitly what is known as ``aliasing'': power in the 
input spectrum at large wavenumbers (i.e. above the sampling frequency) 
gives rise to power at small $k$. Indeed the amplitude
at small $k$ of the discreteness term is proportional to
$I(\mathbf{k} \rightarrow 0)$ in Eq.~(\ref{PS_order1_disc_simplified}), 
which is a sum depending strictly on the power in modes at wave-numbers 
{\it greater than or equal to} the sampling frequency $k_s=2k_N$. 
Further if one cuts at the Nyquist frequency $k_N$, i.e.,
$f(k/k_c)=\Theta(k_N-k)$, where $\Theta$ is the Heaviside step function,
it follows that $I(\mathbf{k})=0$ for $k < k_N$. In this case therefore
we have, for $\mathbf{k}\neq \mathbf{h} $, that
\be
P(\mathbf{k})=P_{\rm th}(k) \Theta (k_N-k) + P^{(1)}_d (\mathbf{k})
\Theta (k-k_N) 
\label{PS-leading-cutkN-final},
\ee 
i.e., to leading order in the input spectrum the full PS of
the generated IC is exactly equal to this input spectrum  
below the Nyquist frequency, and given by the discreteness 
term Eq.~(\ref{PS_order1_disc_simplified}) above the Nyquist frequency. 
It is easy to verify that an analogous
result applies if the cut-off is imposed in the first Brillouin
zone (FBZ), i.e., setting the PS to zero but for vectors with 
all three components $\in [-k_N, k_N]$. In  cosmological simulations 
a cut-off is usually imposed in this way (see e.g. \cite{couchman}, 
\cite{bertschingercode}).

In Fig.~\ref{iofk1} is shown the numerically computed
value of $I(\mathbf{k})$ as a function of $k$, in three 
dimensions\footnote{We show
the average for all vectors $\mathbf{k}$ with modulus in a 
bin centered about $k=|\mathbf{k}|$.}, for a pure power-law PS with $n=-2$, 
(i) without a cut-off [i.e. with $f(k)=1$ in Eq.~(\ref{PS-generalIC})] and, 
(ii) with an abrupt top-hat cut-off at $k_N$, i.e., $f(k)=\Theta(k_N-k)$.
In Fig.~\ref{iofk2}
we show the same quantity but for
$n=0$ and three top-hat cut-offs at $k_N$, $2k_N$, $3k_N$.
We see clearly illustrated the behaviors discussed above.
Note that for $n < -d+2$ ($n >-d+2$) the expression 
for $I(\mathbf{k})$ converges (diverges) without a cut-off,
which explains the choices for the cut-off functions
in the two figures. If a sharp cut-off 
is not implemented at $k_N$ we see that, in all cases,
$I(\mathbf{k})$ is non-zero and approximately constant 
for $k < k_N$. There is thus an associated aliasing 
term which is, to a very good approximation, 
proportional to $k^2$.

\subsection{Accuracy of generation algorithm in $k$ space}

We can now draw clear conclusions about
the accuracy with which the generation algorithm, applied
on a simple lattice, produces a point
distribution with a PS approximating the input PS of the 
form assumed in Eq.~(\ref{PS-generalIC}): 

\begin{itemize}
\item
For $-d<n<4$, and $f$ an abrupt cut-off at $k_N$, we have  
$P(k)=P_{th}(k)$ for $k < k_N$, up to corrections which 
depend parametrically on the
dimensionless quantity $\Delta_{\rm th}^2(k)=k^d P_{\rm th} (k)$.
For $k > k_N$ we have $P(k)=P_{\rm d}^{(1)}(k)$, where the
latter is a discreteness term given explicitly in
Eq.~(\ref{PS_order1_disc_simplified}).

\item If any input power is included above the
Nyquist frequency $k_N$ of the lattice (or, more precisely, 
outside the FBZ of the reciprocal lattice), it leads
to the appearance of power in the IC at $k<k_N$
(i.e. inside the first FBZ). With power included
above the sampling frequency ($k=2k_N$) there is
an aliasing term proportional to $k^2$ at small $k$. 
In this case therefore the range of $n$ which may be 
accurately represented at small $k$ is limited to 
$-d<n<2$.

\item For $n>4$ one always has $P(k) \propto k^4$ at
sufficiently small $k$, and the PS of the point process produced
by the algorithm therefore 
does not approximate the input theoretical spectrum.

\item For $n< -d$ the algorithm is not well defined
because the correlation function of relative displacements
$d_{ij}(\mathbf{r})$ is undefined.  This is true in the 
infinite volume limit. In practice one generates IC 
in a finite system, usually taken to be a cube 
(with periodic boundary conditions). This means that in practice
the input PS is always cut at the corresponding fundamental
frequency of the box, so that, even for $n<-d$, the algorithm
can be applied. The implication of our result is that one will
find in this case that the PS obtained will depend on this box size,
becoming badly defined in the infinite volume limit. We will verify
that this is the case in our numerical study below.

\end{itemize}

\subsection{Glass pre-initial conditions}
\label{Glass pre-initial conditions}

The above conclusions were derived assuming that the  ``pre-initial'' 
distribution is a simple lattice. The alternative starting point 
quite often used in cosmological NBS are ``glassy'' configurations, 
obtained by evolving gravity with a negative sign and a strong 
damping on the velocities \cite{white_leshouches, bertschingercode}. 
Without the damping, this system is essentially
just what is known as the ``one component plasma''  in statistical physics
(for a review, see \cite{OCP-review}).  The small $k$ behavior of the power
spectrum is then expected to be $P_{in}(k) \sim k^2$ at small $k$
\footnote{Here ``small'' means compared to the inverse of 
the Debye length characterising the screening. This statement is
true only if one neglects the damping, and assumes the system 
is in the fluid phase}. With the damping term
what is found is a PS with
a behavior $\sim k^4$ at small scales \cite{smith}. Assuming this 
form for
the spectrum \footnote{We assume thus that 
$P_{\rm in} (k) \sim k^4$ up to  $k$ of order the ``Nyquist
frequency'' (i.e. the inverse of a characteristic interparticle
distance) followed by a flattening to the required asymptotic form 
$P_{\rm in} (k) =1/n_0$ for larger $k$.} it is easy to follow
through the analysis given above for this case. The only change
is that the term $I(\mathbf{k})$ is now
non-zero for all $k$ : because $P_{in} (k)$ is non-zero for
all $k$ it is not possible to have zero overlap of its 
support with that of $P_{\rm th}(k)$ in Eq.~(\ref{PS_order1_disc}).
This  is what permitted this term to be zero in the case
of a lattice and a top-hat cut-off at the Nyquist
frequency. Thus in the case of a glass there will generically
be a correction $\propto k^2$ at $k$ below the wave-number
characteristic of the inter-particle distance in the glass.
Thus the range of power-law spectra which may be accurately
represented by the generation algorithm in this case at 
small $k$ is $-d < n <2$. The models simulated in the context
of cosmological N-body simulations are always well 
inside this range.

What is the source of these limits on the representation
of PS with $n>2$ (or $n>4$ on the lattice)? We remark
that the appearance of such terms \footnote{We note that 
in \cite{melott+shandarin_k4}, which studies an input 
``top-hat'' PS without power at small $k$, the $k^4$ term 
in the PS has actually been actually measured numerically 
in the IC. The authors give it the same physical explanation 
we now discuss.} would appear to be related 
to a well-known argument used by Zeldovich \cite{zeldovich-k4, zeldo}
in determining the limits imposed by causality on fluctuations
(See \cite{peebles_cosmo} and \cite{gabrielli_etal_04} for 
discussion of this result and further
references.): any stochastic process  which moves matter 
in a manner which is correlated only up to a {\it finite} 
scale generates terms proportional to $k^2$ in the PS 
at small $k$.  The coefficient of the $k^2$ term vanishes,
leaving a leading term proportional to $k^4$, if the additional  
condition  is satisfied that the center of mass of the matter distribution 
is conserved (i.e. not displaced) locally
.
The condition on the support of the displacement field required
to make the coefficient of the $k^2$ vanish should thus be equivalent 
to a condition of local center-of-mass conservation under the effect 
of the displacement field.

\section{
Results in real space}
\label{IC_exact_realspace}


We now turn to the consideration of the real
space properties of the distributions generated
by the algorithm. In this section we use the 
$k$ space results of the previous section to
determine these properties approximately, but analytically.
In the next section we will use numerical simulations
in one dimension to show in detail the validity
of these results. 

\subsection{
Definitions and background}
\label{Definitions and background}

The quantities we will study in real space are the
reduced 2-point correlation function $\ti\xi(\mathbf{r})$
and the variance of mass in spheres. In fact we
will principally consider the latter for reasons
which we will explain below.

We recall that $\ti\xi(\mathbf{r})$, for a statistically homogeneous
distribution, is defined by 
\be
\label{definition_xi2}
\lan\overline{\rho(\mathbf r)\rho(\mathbf r')}\ran=\rho_0^2(1+\ti\xi(\mathbf r-\mathbf r')),
\ee
where $\lan...\ran$ is the ensemble average. For a discrete 
distribution (i.e. the case we always consider here) 
$\lan\overline{\rho(\mathbf r)\rho(\mathbf r')}\ran dV_1\,dV_2$ 
is the {\em a priori} probability to find 2 particles in the 
infinitesimal volumes $dV_1$, $dV_2$ respectively around 
$\mathbf r_1$ and $\mathbf r_2$. The correlation function 
$\tilde\xi(\mathbf{r})$ measures therefore the deviation 
of this probability from that in a Poisson distribution
(equal to $\lan \rho_0 \ran ^2 dV_1\,dV_2$). It
is related to the PS as its Fourier transform.

The normalized mass variance $\sigma^2(R)$ in
spheres of radius $R$ is defined as
\be
\si^2(R)=\frac{\langle M(R)^2\rangle-\langle M(R)\rangle^2}{\langle M(R)\rangle^2}.
\ee
where $M(R)$ is the mass in a sphere of radius R, centered
at a randomly chosen point in space. It is given in terms
of the correlation function by
\be
\label{sigma-cfn}
\si^2(R)=\frac{1}{V(R)}\int_{V(R)} d^dr_1 \int_{V(R)} d^d r_2 \ti{\xi}(|\mathbf{r}_1-\mathbf{r}_2|)
\ee
(where $V(R)$ is the volume of a sphere of radius $R$), and in terms of 
the PS by
\be
\label{sigma-PS}
\si^2(R)=\frac{1}{(2\pi)^d}\int d^dk P(\mathbf{k}) |\ti{W}_R(k)|^2
\ee
where $\ti{W}_R(k)$ is the Fourier transform of the window
function for a sphere of radius $R$, normalized so that
$\ti{W}_R(0)=1$.

It is simple to show (see e.g. \cite{peebles_cosmo},
and \cite{glasslike, book} for a more detailed discussion)
that, for a PS of the form 
(\ref{PS-generalIC}), the behavior of the integral
in (\ref{sigma-PS}) depends strongly on the value
of $n$:
\begin{itemize}
\item for $-d< n < 1$ the integral for $\si^2 (R)$
is dominated by modes $k \sim 1/R$ and one
has 
\be
\si^2 (R) \sim k^d P(k) |_{k \sim 1/R}  \propto \frac{1}{R^{d+n}}
\label{variance-n<1}
\ee

\item for $n> 1$ the integral is dominated by modes 
$k \sim k_c^{-1}$ (i.e. by the ultra-violet cut-off) and
one has always 
\be
\si^2 (R) \propto \frac{1}{R^{d+1}}.
\label{variance-n>1}
\ee
\end{itemize}

For $n=1$ one obtains the transition behavior, in which
the integral depends logarithmically on the cut-off $k_c$.
This gives $\si^2 (R) \propto \ln R/R^{d+1}$.

The behavior in  Eq. (\ref{variance-n>1}) is thus actually
a limiting behavior. It is in fact a special case of
a much more general result (see \cite{glasslike, book} for a discussion
and references to the mathematical demonstration of
this result): the most rapid possible decay {\it in any 
mass distribution} of the unnormalized variance of the mass 
$\lan (\Delta M)^2\ran_V$ in a volume $V$ is proportional 
to the {\it surface} of the volume. 

\subsection{Perturbative results in real space}

Returning now to 
Eqs.~(\ref{PS_order0_cont+disc}- \ref{PS_order1_disc}), 
and using  Eq.(\ref{sigma-PS}), we infer that, at 
linear order in the amplitude of the input PS,
we have
\bea
\label{aprox_si}
\si^2(R)&=&\si_{in}^2(R)+\si_{th}^2(R)+\si_{d}^2(R)\\
\label{aprox_xi}
\ti\xi(\mathbf{r})&= &\ti\xi_{in}(\mathbf{r})+\ti\xi_{th}(r)+\ti\xi_{d}(\mathbf{r})
\eea
for the normalized mass variance and correlation function
of the IC. The `in' and `th' subscripts in each case have 
the obvious meanings, with `d' indicating the term
associated to the linear order discreteness correction
$P_d^{(1)}(k)$. We have assumed implicitly that the integrals
pick up negligible contribution from the regions, at
large $k$, where the linear approximation to the full PS
is not good. This will typically translate into a lower bound
on $R$ and $r$ for the validity of 
Eqs.~(\ref{aprox_si}) and (\ref{aprox_xi}).

It is simple to understand from Eqs. (\ref{aprox_si}) and (\ref{aprox_xi})  
why the question of the representation of real space properties of the IC
generated using the ZA is non-trivially different from that of
$k$ space properties. In $k$ space we had analogous expressions
to Eqs. (\ref{aprox_si}) and (\ref{aprox_xi}), from which 
it followed that $P(\mathbf{k}) \approx P_{th}(k)$ to very good
accuracy at small $k$. One necessary ingredient for this was
that the term $P_{in}(k)$ could be neglected at small $k$,
as it is identically zero outside the FBZ on a lattice and
decreasing very rapidly to zero ($\propto k^4$) in a glass.
In real space we do not have the same ``localization'' at
large $k$ of the intrinsic fluctuations associated with
the pre-initial distribution. Indeed we have noted above
that there is a limiting behavior ($\propto 1/R^{d+1}$) to 
the decay with radius $R$  of the mass variance, for
any distribution \footnote{While the result we cited 
concerning the variance applies strictly
to the case of statistically homogeneous and
isotropic distributions, it can be shown 
(see \cite{glasslike, book})
that it applies also to the variance in spheres 
measured in a lattice.}. The amplitude of this 
leading term is fixed by the inter-particle
distance $\ell$, with 
$\sigma_{\rm in}^2 \sim (\ell/R)^{d+1}$, while
that of the two other terms Eqs. (\ref{aprox_si}) 
is proportional to the amplitude $A$ of 
the input spectrum.  Likewise for the correlation
function the intrinsic term $\ti\xi_{in}(r)$ is
generically delocalised in space, and depends 
only on the particle density, while the other 
two terms are proportional to the amplitude
of the input PS. At sufficiently low
amplitude, both quantities will be dominated
at any finite scale by those of the underlying 
pre-initial point distribution, and thus 
will not be approximated by their behaviors
in the input model. This is a behavior which
is qualitatively different to what we have
seen in reciprocal space. We now examine
in a little more detail these two-point quantities.  
We treat them separately as they are quite 
different for what concerns their comparison 
to the continuous theoretical input quantities:
being an integrated quantity, the mass 
variance is intrinsically smooth and can
be directly compared with its counterpart
in the input model. 

\subsection{Mass variance in spheres}
\label{Mass variance in spheres}


Given  Eq.~(\ref{aprox_si}), and the limits
we have discussed on the behavior of
the variance, we can immediately make
a simple classification of the PS of
the form (\ref{PS-generalIC})
for what concerns the representation of
their variance in real space. 
The
faithfulness of such a representation
requires simply
\be
\si_{th}^2(R)\gg \si_{in}^2(R).
\label{condition-dominance}
\ee
For either a lattice or glass we have the
``optimal'' decay  $\si_{in}^2(R) \propto 1/R^{d+1}$.
In order for Eqs.~(\ref{aprox_si}) and
(\ref{aprox_xi}) to be valid 
we require that Eqs.~(\ref{PS_order0_cont+disc}- \ref{PS_order1_disc}) 
be valid. As discussed in the previous section
we expect this to correspond to the criterion that 
$\Delta_{\rm th}^2(k) =k^d P_{th} (k) <1$ for 
the relevant $k$. Given that $P_d^{(1)}(k)$ is at most proportional 
to $k^2$ at small $k$, the associated variance is also 
$\propto 1/R^{d+1}$ above the interparticle
distance $\ell$, and thus sub-dominant with respect to
the leading term at all scales.
Since we generically cut the input spectrum
around $k_N$, and will consider simple power law spectra
up to this scale with $n<-d$, it suffices to have
\be
\Delta_{\rm N}^2\equiv \Delta_{\rm th}^2 (k_{\rm N})= 
k_{\rm N}^d P_{th} (k_{\rm N}) <1 \,.
\label{def-convergence}
\ee
Up to a numerical factor of order unity
this is none other that the criterion
\footnote{For the case $n \geq 1$, this
is true only because the input PS  
is cut at the Nyquist frequency; for
$n<1$ it is true even without the cut-off.}
that $\si_{th}^2(\ell)<1$, and 
so it follows that we expect 
the following behaviors:

\begin{enumerate}

\item For $n >1$ we have seen that $\si_{th}^2(R)\sim 1/R^{d+1}$,
i.e., $\si_{th}^2(R)$ has the same functional behavior as that of the
``pre-initial'' variance. Given that the former is necessarily
smaller at the inter-particle distance, the condition 
Eq. (\ref{condition-dominance}) will never be fulfilled,
as the full variance will be dominated by that
of the pre-initial configuration. 

\item For $-d<n<1$ we have that 
$\si_{th}^2(R)\sim 1/R^{d+n}$, 
which thus decays more slowly than the 
``pre-initial'' term. Thus there will be a 
scale $R_{min}$ such
that for $R> R_{min}$ one can
satisfy the condition Eq. (\ref{condition-dominance}).
It is easy to infer that, for any $d$, we have  
\be
\label{condition-variance-n<1}
R_{min} \sim \ell \left(\frac{1}{\Delta_{\rm N}}\right)^{\frac{2}{1-n}}
\ee

\end{enumerate}

\subsection{Two point correlation function}
\label{Two point correlation function}

The case of the two point correlation function is similar.
The determination of the range of faithful representation 
of the theoretical correlation function is, however, 
more complicated by the very non-monotonic behavior
of the correlation function in both the (unperturbed) 
lattice and glass. This leads, as we will explain, to
a strong dependence on how the correlation function is smoothed
when it is estimated in a sample.

Unlike for the variance, there is no intrinsic limit on the 
rapidity of the decay of the correlation function for point 
processes. Indeed for a Poisson process one has 
$\ti\xi_{in}(r)=0$ for $r>0$, and exponentially decaying
correlation functions are commonplace in many physical 
systems. For both a lattice and glass distributions the
leading term $\ti\xi_{in}(r)$ in Eq.~(\ref{aprox_xi}) presents a very
non-trivial behavior. The two point correlation function of
the lattice is in fact not a function of $r$, but a distribution
which depends on $\mathbf{r}$: it is proportional to a Dirac delta 
function when $\mathbf{r}$ links any two lattice points, and 
equal to $-1$ otherwise (see Appendix \ref{1d} for the 
explicit expression). For the glass the correlation function
is not known exactly --- it depends on the details of the generation
of the glass configuration used --- but generically it will be 
expected to have a similar oscillatory structure describing its very 
ordered nature, with decay only at scales considerably above
the interparticle distance \footnote{The characteristic property
of these configurations is that the force on particles is extremely
small. This imposes a very strong correlation between the
positions of particles. In studies of the one component
plasma, mentioned above \cite{OCP-review}, the appearance of multiple 
peaks in the correlation function is observed
as the temperature is lowered.}.  This underlying highly ordered 
structure is evidently not washed out by the application of very 
small displacements. In particular for relative displacements
much smaller than the initial interparticle separation, it
is clear that the form of the underlying correlation function
will remain highly oscillatory up to a scale considerably
larger than the interparticle distance. Just as in the case of the
mass variance, therefore, one can conclude that the theoretical
term in Eq.~(\ref{aprox_xi}) will always be dominated by the 
discreteness terms up to some scale, which becomes larger as 
the input amplitude is decreased.  

A simple analytical estimate, like that given above for the variance,
of the scale at which the theoretical term will dominate the 
discreteness terms, and thus at which the input theoretical
two-point correlation function is well approximated by that of 
the generated IC, is not possible: for the lattice such an
estimate must take also into account the term $\ti\xi_{d}(\mathbf{r})$
which together with  $\ti\xi_{in}(\mathbf{r})$ gives a regular 
oscillating and decaying function; for a glass we do not
have the analytical form of the correlation function.

There is a further important difficulty if one wishes to
compare the correlation function in generated IC with 
the input one. In estimating the correlation function in a finite
sample one must introduce a finite smoothing: one computes 
it by counting the number of pairs of points with separations 
in some finite interval, typically a radial shell of some
chosen thickness. Indeed while the full correlation function 
is in general a function of $\mathbf{r}$, this procedure makes it 
a function of $r$ like the theoretical correlation function.
Given that, at low amplitude of the relative displacements,
$\ti\xi (\mathbf{r})$ has both a strongly oscillating and
strongly orientation dependent behavior, such a smoothing
can change very significantly its behavior. Thus the
scale at which agreement may be observed between the measured 
ensemble averaged two point correlation function and that of
the input model will depend both on $A$, $\ell$ and the 
precise algorithm of estimation of the correlation function.

\section{
Numerical study in one dimension}
\label{comparison}

In this section we study the generation algorithm
using numerical simulations. This allows us 
to verify our conclusions about two point properties
in reciprocal and real space, derived in the limit
of small amplitudes of the input PS. Further it allows
it to show the accuracy of the full analytic expression 
Eq.~(\ref{P_3d}), for any input amplitude. We work in 
one dimension because of the numerical feasibility of the study
in this case: we measure directly the real-space 
mass variance for a large ensemble of configurations, which 
is not numerically feasible (for modest computational 
power) in three dimensions.  The exact ensemble 
average results given above, on the other hand, are
easily calculated. The simplified and more explicit 
expressions for the relevant quantities are given in
Appendix~\ref{1d}.  There is no intrinsic
difference of importance between one and three
dimensions for the questions we address\footnote{One
minor exception for the case of the two point
correlation function, related to the last point 
discussed in the previous section, is discussed at 
the appropriate point below.}.

We consider the case in which the
pre-initial distribution is a lattice.
Following our discussion in the previous sections
we study separately the four following specific
examples for input PS as 
in Eq.~(\ref{PS-generalIC}): (i) n=-1/2 
(example of $-d<n<1$), (ii) n=3 
(example of $1<n<4$), (iii) n=7
(example of $n>4$) and (iv) n=-2
(example of $n<-d$, in which case we
have found the algorithm to be badly defined in the
infinite volume limit). We will specify
the cut-off function in each case. We then
also present numerical results for the two point
correlation function in just the
first of these models to illustrate the 
discussion of this quantity given at the end
of the preceding section.

\subsection{$n=-1/2$ (Case $-d<n<1$)}

\begin{figure}
\resizebox{8cm}{!}{\includegraphics*{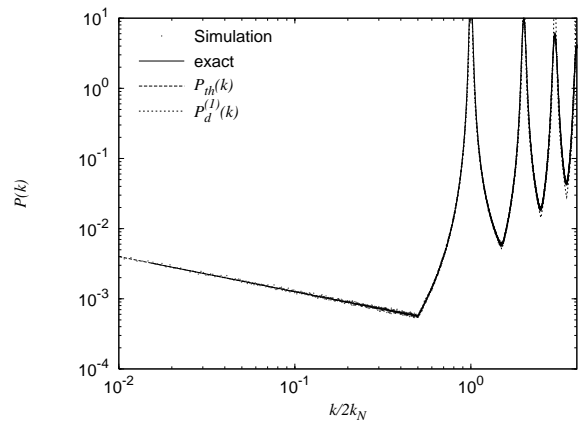}}
\caption{PS of a model with $n=-1/2$, sharp cut-off $f(k)=\Theta(k-k_N)$
 and $\Delta_N=1.77\times 10^{-3}$ ($A=10^{-3}$).  The simulation
 results are averaged over one thousand realisations of IC, generated using
 the standard algorithm (adapted to one dimension).
\label{Fig1}}
\end{figure}

\begin{figure}
\resizebox{8cm}{!}{\includegraphics*{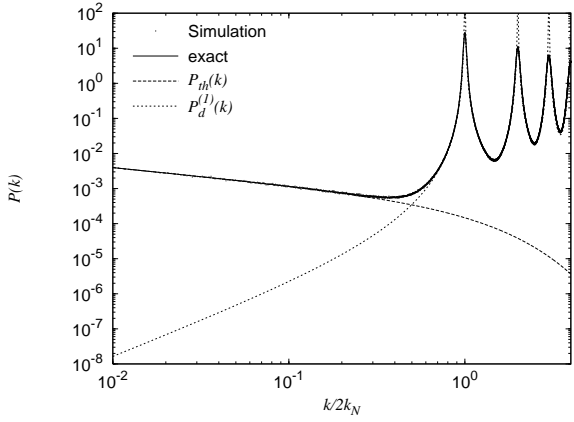}}
\caption{PS of a model with $n=-1/2$, continuous cut-off
 $f(k)=\exp(-k/2k_N)$ and $\Delta_N=1.77\times 10^{-3}$ ($A=10^{-3}$).
\label{Fig2}}
\end{figure}

In Fig.~\ref{Fig1} are shown results for an input PS 
$P_{th} (k)=Ak^{-1/2}$ with $A=10^{-3}$, which corresponds
to $\Delta_N=1.77\times 10^{-3}$. Here, as in the rest of
this section, we use {\it units of length 
in which the inter-particle distance is equal
to unity}.  We have imposed a sharp FBZ 
cut-off $f(k)=\Theta(k-k_N)$. In the figure we see,
as expected, excellent agreement between the 
PS measured by averaging over a thousand
realisations of IC, generated using
the standard algorithm (with Gaussian displacements)
in a periodic interval containing a thousand particles,
and the theoretical expression at linear order, as
given in the previous section. Note that on the $x-axis$
is given $k/2k_N$, so that first Bragg peak appears at
unity, and the sharp change in the PS at $0.5$. 

Fig.~\ref{Fig2} differs only in that we have now imposed 
a continuous cut-off $f=e^{-k/2k_N}$. Again we observe, as expected,
excellent agreement between the measured PS and the
theoretical expression. The agreement between the input
PS and the measured PS is, however, less perfect around 
$k_N$, because the discreteness term $P_d^{(1)} (k)$
contributes now inside the FBZ (i.e. for $k < k_N$).
The effect is, however, very small as the latter term
is, in this range, proportional to $k^2$.

\begin{figure}
\resizebox{8cm}{!}{\includegraphics*{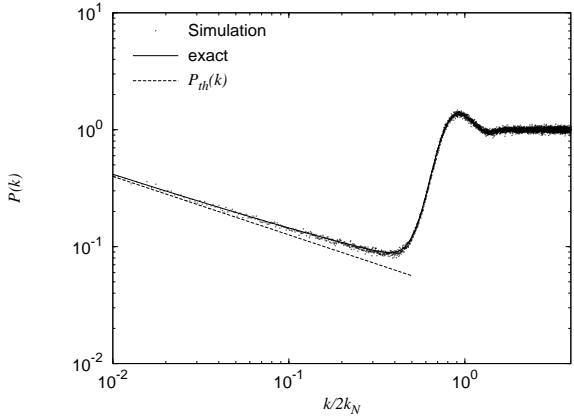}}
\caption{PS of a model with $n=-1/2$, sharp cut-off $f(k)=\Theta(k-k_N)$
 and $\Delta_N=0.18$ ($A=0.1$).
\label{Fig3}}
\end{figure}

In Fig.~\ref{Fig3} are shown results for the same shape
PS, but now with a higher amplitude, $A=0.1$, corresponding
to $\Delta_N=0.18$. The cut-off here is sharp. Shown are 
the input theoretical PS, the average over one thousand
realisations, and the exact expression for the PS. We are 
not in this case in the regime in which the perturbative
expansion of the full PS is valid at $k_N$, and therefore 
do not plot $P_d^{(1)}(k)$ as in the previous figures. Indeed 
we see that the PS of the generated IC begin to deviate
sensibly from the input theoretical IC already at a $k$ 
significantly smaller than $k_N$, with a discrepancy of 
about a factor of two in the amplitude at $k=k_N$.
Note that, nevertheless, the results of the simulations
agree extremely well with the exact expressions for the
full PS.

\begin{figure}
\resizebox{8cm}{!}{\includegraphics*{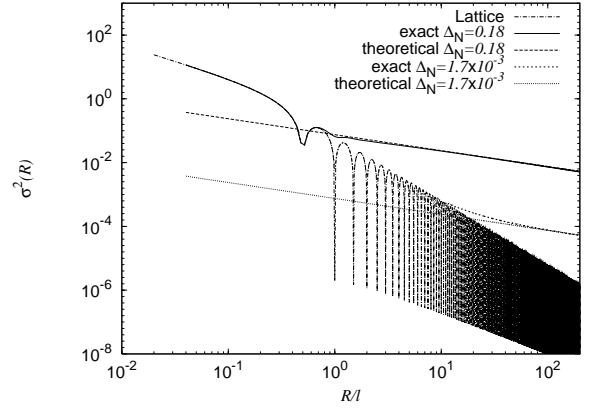}}
\caption{Mass variance in spheres of a model with $n=-1/2$, sharp
 cut-off $f(k)=\Theta(k-k_N)$  and two different amplitudes of the
 theoretical PS. 
\label{Fig10}}
\end{figure}

In Fig.~\ref{Fig10} are shown the 
real-space variance in spheres of radius $R$ 
(i.e. intervals of length $2R$) for the case of the
sharp cut-off and the two different amplitudes just 
considered. The curves labelled ``exact'' are
those corresponding to the ensemble average of the 
full IC, and those labelled ``theoretical'' are
those of the input model. We see clearly illustrated
the results anticipated in the previous section:
for low amplitudes the exact curve is dominated 
at small distances by the variance of the underlying
lattice, and the low amplitude theoretical 
expression (which has a behavior 
$\sigma^2(R) \propto 1/R^{1/2}$) is approximated 
only once this term coming from the lattice
(with $\sigma^2(R) \propto 1/R^{2}$) has decayed
sufficiently. At the higher amplitude the theoretical
expression, on the other hand, is well approximated 
for scales just above the interparticle distance
\footnote{The discrepancy between the variance appears
smaller than that in the PS (shown in Fig.~\ref{Fig3} )
at the inverse scale due to the different 
range of scale on the y-axis in the two plots.
The relative difference is in fact of the same order.}. 
 
\subsection{ $n=3$ (Case $1<n<4$) }

\begin{figure}
\resizebox{8cm}{!}{\includegraphics*{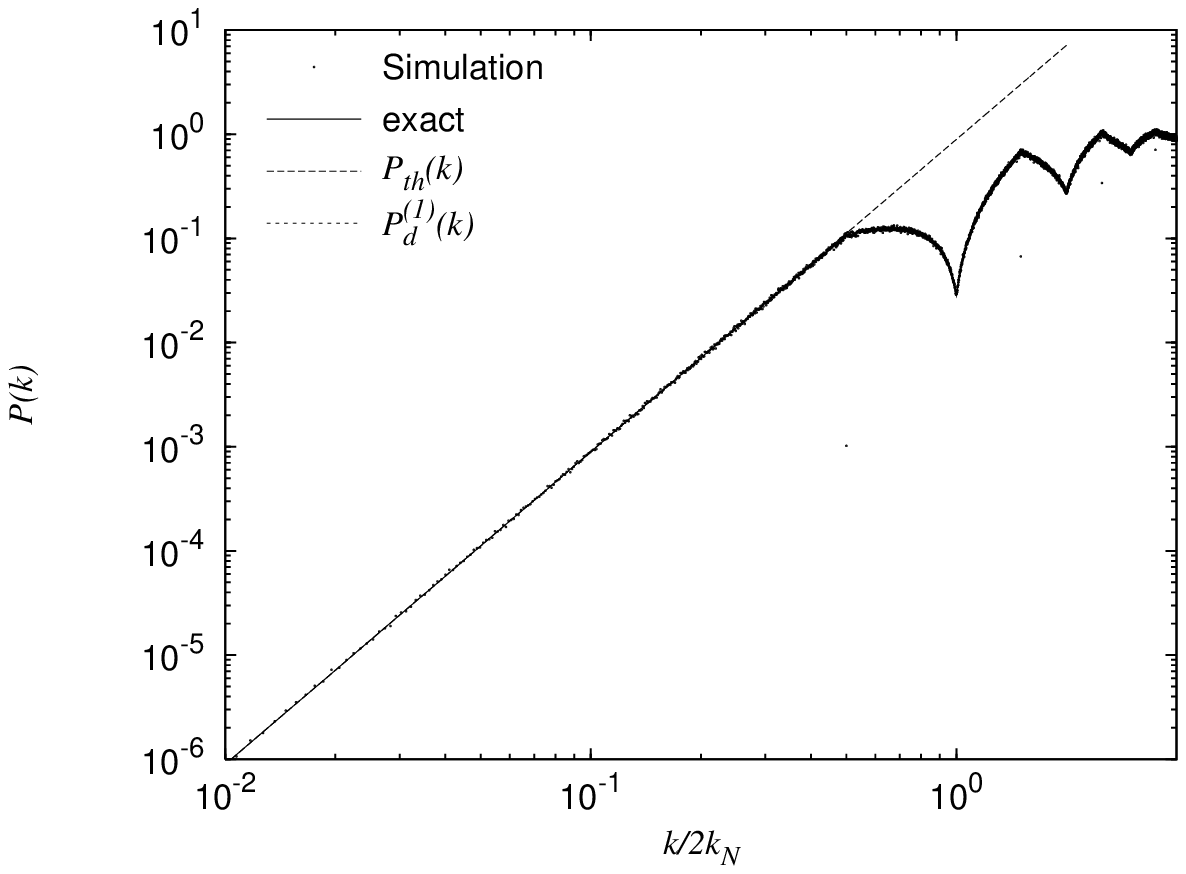}}
\caption{PS of a model with $n=3$, sharp cut-off $f(k)=\Theta(k-k_N)$ and $\Delta_N=0.35$ ($A=1.8\times10^{-5}$).
\label{Fig4}}
\end{figure}

\begin{figure}
\resizebox{8cm}{!}{\includegraphics*{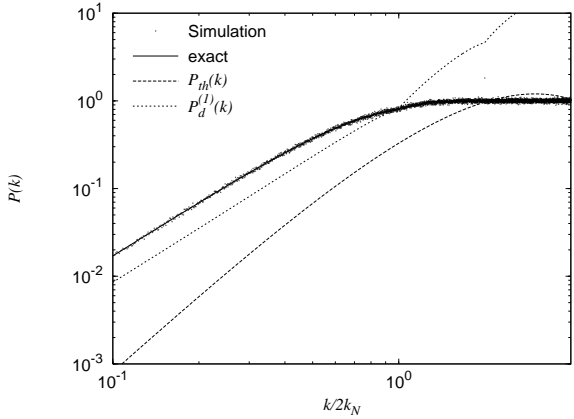}}
\caption{PS of a model with $n=3$, continuous cut-off $f(k)=\exp(-k/2k_N)$ and $\Delta_N=0.35$ ($A=3.6\times10^{-3}$).
\label{Fig5}}
\end{figure}

\begin{figure}
\resizebox{8cm}{!}{\includegraphics*{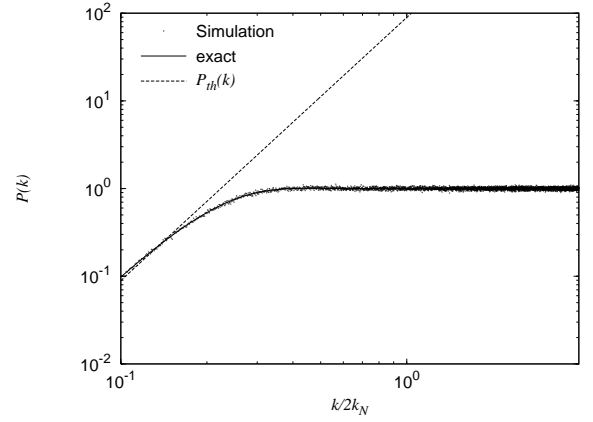}}
\caption{PS of a model with $n=3$, sharp cut-off $f(k)=\Theta(k-k_N)$ and $\Delta_N=35$ ($A=0.36$).
\label{Fig6}}
\end{figure}

\begin{figure}
\resizebox{8cm}{!}{\includegraphics*{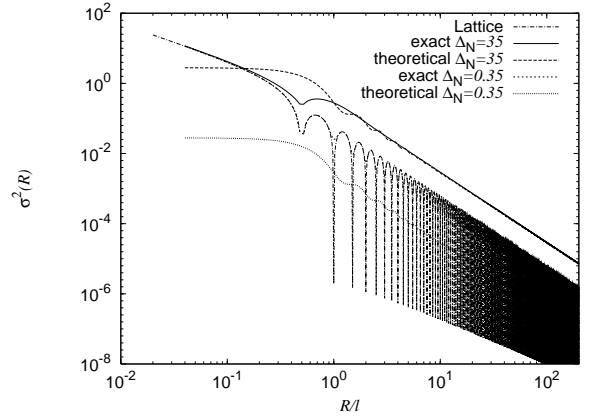}}
\caption{Mass variance in spheres of a model with $n=3$, sharp cut-off $f(k)=\Theta(k-k_N)$ and and two different amplitudes of the theoretical PS.
\label{Fig11}}
\end{figure}

Figs.~\ref{Fig4} to ~\ref{Fig11} show exactly the same quantities
as the four previous figures, but now for an input power-law
PS  with $n=3$. The two amplitudes chosen are given in the captions,
the low amplitude corresponding to the case where the linear 
approximation to the exact formula for the PS is a good approximation.
Figs.~\ref{Fig4} to ~\ref{Fig5} illustrate the more important difference
that arises in the case that $n>2$ when the cut-off imposed on the PS is 
smooth instead of being imposed sharply inside the FBZ: the $k^2$
term at small $k$ generated in $P_d^{(1)}(k)$ in this case
dominates the input PS at small $k$ so that it is no longer
faithfully represented by the PS of the generated IC at any $k$.
Fig.~\ref{Fig6} shows essentially the same thing as 
Fig.~\ref{Fig3}. For higher amplitudes the agreement of the
input PS with that of the generated IC is shifted to smaller
$k$. The exact formula for the PS agrees very well with that
of the generated IC measured in the simulations, but the
linear approximation to the discreteness effects at larger
$k$, given by $P_d^{(1)}(k)$, is no longer a good approximation.

Comparison of Fig.~\ref{Fig11} with Fig.~\ref{Fig10} shows 
the difference between the cases $n<1$ and $n>1$ for what 
concerns the behavior of the mass variance in real space. 
Because the theoretical variance has the same scale dependence
($\sigma^2(R) \propto 1/R^{2}$) as the lattice variance, the
latter always dominates the former if the amplitude is low.
Specifically, if the input mass variance at the lattice
spacing is less than that of the lattice (which is of order
unity) the mass variance of the IC is not approximated
at any scale by that of the input model.

\subsection{$n=7$ (Case $n>4$)}

\begin{figure}
\resizebox{8cm}{!}{\includegraphics*{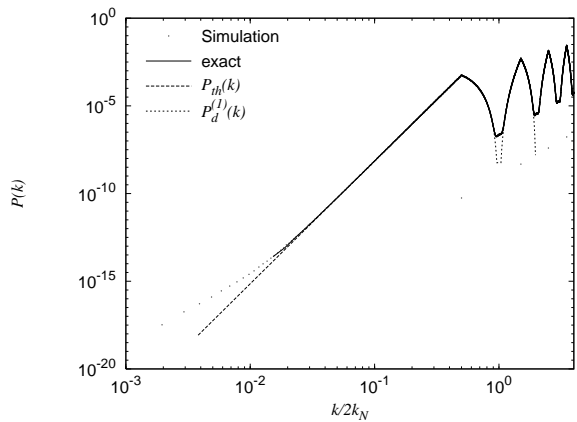}}
\caption{PS of a model with $n=7$, sharp cut-off $f(k)=\Theta(k-k_N)$ and
$\Delta_{\rm N}=1.76\times 10^{-3}$ ($A=1.85\times10^{-5}$).
\label{Fig7}}
\end{figure}

\begin{figure}
\resizebox{8cm}{!}{\includegraphics*{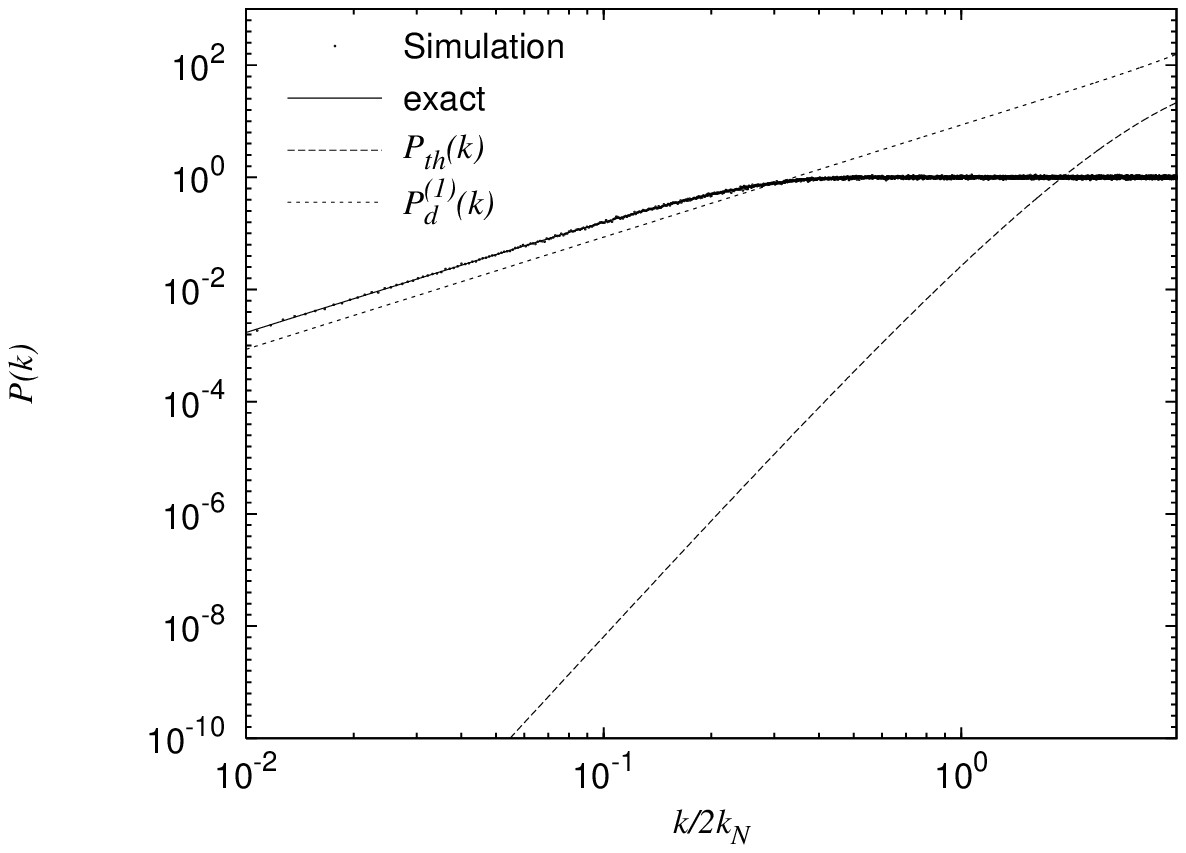}}
\caption{PS of a model with $n=7$, continuous cut-off $f(k)=\exp(-k/2k_N)$ and
$\Delta_N=1.76\times 10^{-3}$ ($A=4.8\times10^{-5}$). 
\label{Fig8}}
\end{figure}

Figs.~\ref{Fig7} to ~\ref{Fig8} show results for the PS of a single
low amplitude $n=7$ input model, for the case of a sharp and continuous
cut-off respectively. These figures illustrate the limitation discussed
in the previous section for the representation of a small $k$ input
PS with $n>4$. Using the sharp cut-off inside the FBZ 
the term $P_d^{(1)}(k)$ is zero for $k$ inside the FBZ, but
nevertheless the theoretical behavior at small $k$ is not
represented because the corrections to
Eq.(~\ref{PS-leading-cutkN-final}), at quadratic order in
the amplitude $A$, are non-zero. Thus at asymptotically small
$k$ we see the PS of the generated IC deviate from the input one
\footnote{The numerical integration of the exact expression
in this case is very difficult because of a very rapidly oscillating 
behaviour in $d(x)$ at large $x$. The `exact' curve has 
thus been calculated just far enough at small $k$ so 
that the deviation from the input PS may be discerned.}. 
Further the behavior at the smallest $k$ is well fit by 
a $k^4$ behavior, which is shown in Appendix \ref{expansion-Pc} 
to be that of the quadratic order correction. 

In Fig.~\ref{Fig8} one can observe the dramatic effect,
as we saw illustrated also in Fig.~\ref{Fig6}, of using a 
continuous cut-off for the case $n>2$. Just as in the
case of $n=3$ we see that the input PS is no longer 
well approximated --- indeed not even poorly approximated ---
by the PS of the generated IC. Note that, differently
from Fig.~\ref{Fig7}, there is no range of 
intermediate $k$ where the input PS is approximated.
This is because it is the correction $P_d^{(1)}(k)$
which dominates at small $k$, with an amplitude 
proportional to same (linear) power of the input
PS. Correspondingly in Fig.~\ref{Fig7} the $k$
at which a deviation towards the $k^4$ behavior is
observed can be shifted to arbitrarily small $k$
by taking a sufficiently low initial amplitude.

\subsection{$n=-2$ ($n <-d$)}

\begin{figure}
\resizebox{8cm}{!}{\includegraphics*{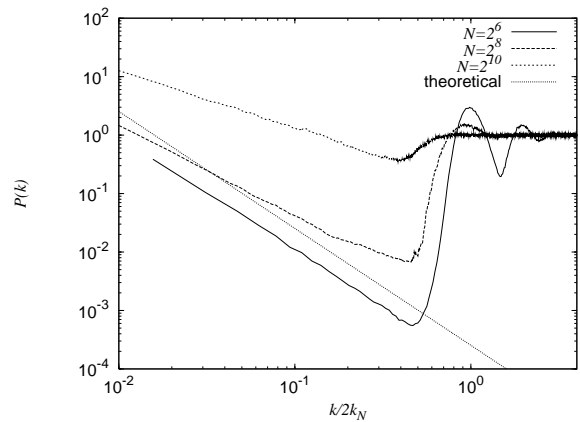}}
\caption{PS from simulations of a model with $n=-2$, sharp cut-off 
$f(k)=\Theta(k-k_N)$ and
$\Delta_N=2\times 10^{-3}$ ($A=4.8\times10^{-2}$), for different number of particles.
\label{Fig9}}
\end{figure}

We show finally in Fig.~\ref{Fig9} results for the PS for
averages over simulations of the case $n=-2$.  In this case, as 
we have discussed above, the algorithm is not well defined in the 
infinite volume limit, because the variance of relative 
displacements at any scale is a divergent. The implementation
of the algorithm in a finite sample, with periodic boundary
conditions, is perfectly well defined as the spectrum of modes
is cut-off at small $k$ by the fundamental, fixed by the 
box size. In the figure we show the results for the PS of
the averages of $1000$ generated configurations, for 
different numbers of particles, i.e., for different
sizes of the system. As anticipated the results depend
strongly on the box size, and neither the amplitude nor
the shape of the input PS is approximated well by
that of the generated distributions.

\subsection{Two point correlation function}

\begin{figure}
\resizebox{8cm}{!}{\includegraphics*{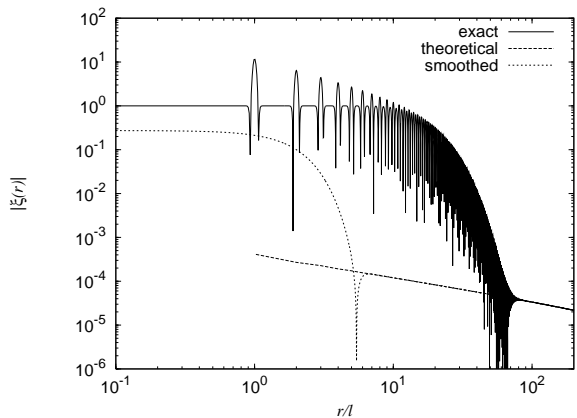}}
\caption{
The absolute value of the measured and theoretical 
two point correlation function for a theoretical input model with 
power-law PS and $n=-1/2$, for amplitude $A=2\times 10^{-3}$.
The curve labelled `exact' is the result of a numerical
evaluation of the full expression of the correlation function
for the given model. The curve `smoothed' gives the same quantity, 
but now smoothed by a convolution with a Gaussian window 
function as given in Eq.~(\ref{smoothing_xi}) with
$W_L(k)=e^{-k^2}$. The `theoretical' curve is the correlation
function of the input model, proportional to $1/r^{1/2}$.
\label{xi_critb}}
\end{figure}

\begin{figure}
\resizebox{8cm}{!}{\includegraphics*{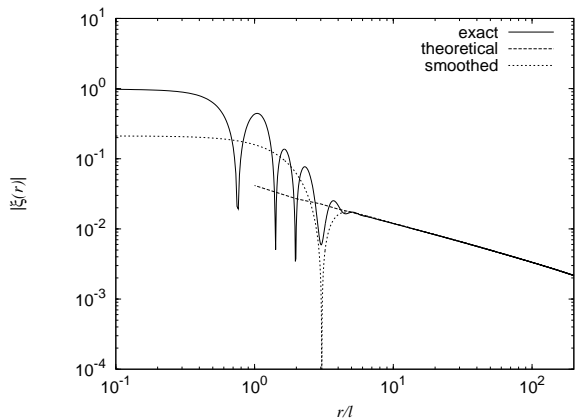}}
\caption{The same quantities as in Fig.~\ref{xi_critb}
for a much larger amplitude, $A=0.2$ of the input PS.
The smoothing is exactly the same as in the previous figure.
\label{xi_crit}}
\end{figure}
 
Figs.~\ref{xi_critb} and \ref{xi_crit} illustrate quantitatively
the discussion and conclusions in Sec.~\ref{Two point correlation function}
above. They show both the exact two point correlation function,
and a smoothing of it, for IC corresponding to an input power-law 
PS with $n=-1/2$.  The smoothing is defined by a convolution
of the discrete density distribution with a spatial window 
function $W_L$:
\be
\rho_c(\br)=\int_{-\infty}^{+\infty}d^3r ^\prime W_L(|\br - \br ^\prime|)
\rho_d(\br ^\prime),
\label{convolution-rho}
\ee
where $\rho_c(\br)$ is the density function of the continuous 
field, $\rho_d(\br)$ of the discrete distribution and $L$ is 
the characteristic scale introduced by 
the smoothing. For the correlation function this gives
\be
\label{smoothing_xi}
\tilde\xi_{s}(\br)=
\int_{-\infty}^{+\infty}dr^\prime
\mathrm{FT}_{\br-\br^\prime}\left[|W_L(k)|^2\right] \tilde\xi (x'),
\ee
where FT denotes the inverse Fourier transform of $W_L(k)$.
For the latter we have taken here a simple Gaussian form
as specified in the caption of the figures.

We observe that the range of agreement between
these quantities and the theoretical correlation function is different 
--- illustrating that the result depends on the smoothing  --- 
and, further, that this range depends also on the amplitude of
the input model. Just as for the mass variance, the scale above
which the theoretical and measured quantities converge increases
(for any given method of estimation/smoothing) as the amplitude
decreases. Further for sufficiently low amplitude perturbations 
the underlying structure of the lattice becomes visible if the
estimated two point function is resolved to the required level
(by a sufficiently narrow smoothing).

One remark is appropriate here on the relation between these
results and those in three dimensions. One important effect 
in that case is not illustrated by these results: if one
takes, as is usually done, a simple pair estimator for $\tilde{\xi}(r)$
using spherical shells of equal width, the volume of the shells grows
as $r^2$. Therefore the oscillations of the true lattice or glass
correlation function will be attenuated much more rapidly as a function
of distance than by the smoothing considered here in one dimension. 
This, however, does not 
change any of the conclusions above: the scale at which agreement
will be observed between the measured and theoretical quantities
will depend on the size of the bins, and taking sufficiently small
bins one can always make the oscillatory structure 
of the underlying correlation function dominate for
a sufficiently low amplitude of the input spectrum.

\section{Summary and Conclusions} 

We first summarize our findings on the accuracy and limitations
of the standard algorithm for generating IC for cosmological 
simulations. We then discuss the conclusions we can draw, in the 
light of our analysis, about
the some numerical work on IC \cite{Baertschiger:2001eu, alvaro_knebe2} 
which partly motivated our study. Finally we turn to the relevance of
our results to the problem of understanding discreteness effects in
the evolution of cosmological simulations. 

\subsection{Results on generation algorithm}
 
We have investigated systematically the algorithm used to 
generate IC of N-body simulations in cosmology, for {\it any} 
given input PS. More specifically we have focussed on the comparison 
of the two point correlation properties, in real and reciprocal space, 
of the IC with those of the input theoretical models. We
consider input PS which are a simple power-law $P(k) \propto k^n$,
but the corresponding results for more complicated cases may
be easily inferred. Our main results are:

\begin{enumerate}
\item Applied on a grid with appropriate sharp cut-off at the
Nyquist frequency $k_N$, the point distribution produced by the
algorithm has PS exactly equal to the input one, below $k_N$,
to linear order in the amplitude and for $-d<n<4$. For $k>k_N$ 
we have also given the exact expression for the PS, which is 
thus the leading discreteness correction in this space. It is 
a term of high amplitude, with a damped oscillating form 
with maxima at the Bragg peaks of the underlying lattice.

\item Applied to a `glass' pre-initial configuration, the
result is almost the same, except that the discreteness
correction has a small $k$ tail proportional to $k^2$.
Thus the range of ``faithful representation'' of the PS
is $-d<n<2$. This latter restriction is not of relevance
to current cosmological models, for which the effective
exponent at all $k$ is within this range.

\item  The algorithm does not produce IC representing 
faithfully an input PS with $n>4$ for arbitrarily 
small $k$. There is the case because there is a term 
proportional to $k^4$ in the PS of the generated PS,
at second order in the amplitude of the input PS.

\item For the case $n<-d$ the algorithm is not
well defined in the infinite volume limit, and we 
have verified that results in a finite volume depend 
strongly on the volume.

\item The transposition of these results to real
space is more subtle than one might have anticipated,
due to the fact that the mass variance and two point
correlation of the underlying `pre-initial' point
distribution are delocalised in this space.

\item For models with $-d<n<1$ the real space variance
in spheres
can be well represented by the generated configurations
starting from a finite scale $R_{min}$ proportional to
the inter-particle spacing. For typical chosen input 
amplitudes it is a few times this distance, but we note
that it diverges as the amplitude of the input model 
goes to zero. 

\item For models with $n>1$, the real space variance
is always dominated, at linear order in the amplitude,
by the ``pre-initial'' variance of the lattice or glass.

\item The conclusions concerning the representation of
the reduced two point correlation function are quite 
similar to those for the mass variance: the theoretical
properties are recovered above a finite scale proportional
to the inter-particle distance, which diverges as the 
amplitude goes to zero.  In practice there is a further
difference with respect to the mass variance, in that
the value of this scale depends also on the smoothing 
is necessarily introduced in estimation of the correlation 
function. For a sufficiently narrow smoothing the correlation 
function will always show at a given scale, for sufficiently
low amplitude of the input model, the underlying 
structure of the lattice or glass configuration.

\end{enumerate}

\subsection{Comments on precedent literature }

Let us now consider, in the light of these results, the
articles \cite{Baertschiger:2001eu, alvaro_knebe, Baertschiger:2003jt,
alvaro_knebe2} which have partly motivated this work. These 
two collaborations draw, on the basis of numerical studies, very
different conclusions about the measured mass variance in spheres 
and two-point correlation function of the IC of cosmological
NBS. 

In cosmology the IC of NBS are invariably studied only in reciprocal
space, simply because it is the natural one for the description of
cosmological models at early times.  In the first of these papers 
\cite{Baertschiger:2001eu} the authors examined instead IC in real 
space, through a numerical study of the IC of some large cosmological 
simulations performed by the Virgo consortium \cite{virgo}. Their 
finding was, very surprisingly, that the measured and theoretical 
values of both the mass variance in spheres and the two point 
correlation function did not match. In \cite{alvaro_knebe} the same 
analysis was repeated by a different set of authors, and an error in the 
normalization in \cite{Baertschiger:2001eu} of the theoretical variance 
was identified. Correcting for this error the authors concluded that 
the agreement between the measured and theoretical properties was good 
for the variance, while the authors of \cite{Baertschiger:2001eu}, in a 
reply \cite{Baertschiger:2003jt}, argued that the agreement was still 
very poor. For the two point correlation function the results of both
sets of authors agreed, showing an estimated correlation function 
qualitatively and  quantitatively different to the expected one. 
The two sets of authors gave a quite different
interpretation to this discrepancy: in \cite{Baertschiger:2001eu} 
it was attributed to a probable systematic difference
between the two quantities due to the underlying correlation
in the ``pre-initial'' configuration, while \cite{alvaro_knebe} 
argued that it was more likely simply due to statistical noise
of the estimator. In a further article \cite{alvaro_knebe2} the second 
authors analyzed these same quantities in the IC of another
set  of cosmological simulations, and arrive at the same 
conclusions as in \cite{alvaro_knebe} concerning both quantities.

For what concerns the mass variance we have seen that the degree of 
agreement between the theoretical and measured variance depends
on the normalization of the model, i.e., on the initial red-shift of 
the simulation. Neither collaboration has studied the dependence of 
their conclusions on this crucial parameter, nor identified it as 
relevant. Thus the conclusions of \cite{alvaro_knebe2, alvaro_knebe} 
about the reliability in general of the representation of the input 
mass variance by the IC are, strictly, incorrect. However, their 
conclusion that the representation of this quantity is good for 
the specific set of IC considered --- normalized at an amplitude
which is typical in practice in cosmological simulations -- is
correct. That is the agreement they observe in a modest range
(see e.g. the figure 3 in \cite{alvaro_knebe2}), from a few times 
the interparticle distance to a scale approaching the box size,
at which finite size effects start to play a role, is real
(rather than purely accidental as is implicitly suggested by 
\cite{Baertschiger:2001eu, Baertschiger:2003jt}). However the
dominant lattice term can clearly be identified at smaller 
scales, and it is evident in view of our discussion that
the range of agreement will decrease (and ultimately disappear)
if one considers the same model with a lower normalization.

For the two point correlation function,
we have seen that the degree of agreement depends not
only on the amplitude of the input model, but also on 
the details of the spatial smoothing in the estimator. 
Again neither collaboration has pinpointed explicitly
the importance of this consideration in evaluating 
the faithfulness of the representation. The authors of
\cite{Baertschiger:2001eu, Baertschiger:2003jt}), however,
are correct when they argue that the difference observed
is a systematic one, and that the oscillating behavior
observed is due to the correlations in the underlying 
pre-initial (lattice or glass) configuration. In
attributing the difference to ``noise'' the  other group
is incorrect, insofar as such noise would be a finite
size effect which should disappear in
the ensemble average. However, noise can of course 
play a crucial role in a finite sample in hiding the
underlying signal in the regime in which it may, in
principle, approximate well the theoretical model. 

\subsection{Physical relevance of results on IC}

We have considered in this paper solely the question of the 
accuracy with which the standard algorithm for generating
IC for cosmological NBS represents the theoretical correlation 
properties. This question is essentially interesting insofar as 
it is relevant to
the question addressed by the series of articles of which
this is the first: the quantification of the differences
between the results of {\it evolved} N body simulations and the 
corresponding theoretical quantities. This question will
be addressed fully in the subsequent papers, and we limit
ourselves now to a partial discussion of the physical 
relevance of our findings. 

The most important result from a practical point of view
is that, at linear order in the theoretical density perturbations,
there is a contribution to the PS of the IC additional to
the theoretical PS. This is a source for gravitational structure
formation through the Poisson equation, which in a given
simulation cannot be separated from the theoretical term.
Indeed we note that the linearity of this term in the 
amplitude of the relative displacements means that, if
the early time evolution follows the Zeldovich approximation,
this term is amplified linearly, just like the theoretical
term. On the other hand, it contributes significantly
only above the Nyquist frequency, and therefore, given
that gravity tends to transfer power very efficiently from
large to small scales (see, e.g., \cite{Little+weinberg+park_1991}), 
one would expect its effects to be washed out over time. However 
if one wishes to quantify precisely discreteness effects, our 
quantification of this leading discreteness contribution in the 
IC is an important first step.

In quantifying such effects it is important also to
first understand the recovery of the continuum limit.
Our results here, as we will now discuss, actually
are quite informative in this respect. Let us consider
the limit in which one recovers {\it exactly} the properties 
of the theoretical continuum model. Given an input theoretical 
model for a cosmological NBS, we introduce two parameters 
with the standard method of discretization we have discussed
here~\footnote{In reality there is of course
also the box size $L$, which we have taken in our study
to be infinite. The finite particle number $N$ is given
by $(L/\ell)^3$.}: $\ell$, the 
lattice spacing in physical units, and the initial
red-shift $z_i$ (which fixes the amplitude $A$ of 
the input PS, with $A \rightarrow 0$ as 
$z_{\rm i} \rightarrow \infty$). 

The continuum limit should evidently correspond to taking
$\ell \rightarrow 0$ (and thus $k_N \rightarrow \infty$).
Let us consider first taking $\ell \rightarrow 0$ 
at fixed $z_{\rm i}$. This corresponds in our analysis
above to working at fixed amplitude of the PS. 
Our results above tell us that the representation of the
PS is good provided we satisfy the condition 
Eq.~(\ref{validity-criterion-2}) for the validity
of the perturbative expansion. This quantity in fact 
converges to zero for $k_N \gg k_c$, and so the
criterion for good agreement in $k$ space for
all $k$ is simply $\Delta_{\rm th}^2 (k_c) \ll 1$.
This agreement becomes arbitrarily good as we
take $z_{\rm i} \rightarrow \infty$ (i.e. 
$z_{\rm i} \rightarrow 0$ ).
Likewise in real space, it follows from 
Eq.~(\ref{condition-variance-n<1}) that 
we converge towards an arbitrarily good
representation of the mass variance when
the limit is taken in this way. The same is
true of the two point correlation function.
Thus the correlation properties of the discretized  
IC converge exactly to the continuum IC.

Our results concerning the differences in 
real space quantities concern the limit 
$z_{\rm i} \rightarrow \infty$, at 
fixed $\ell$. We have seen that there is,
in this case, no convergence towards the 
continuum model. Thus, in the IC, the order of 
the limits in $\ell$ and $z_{\rm i}$ cannot
be interchanged. It will be shown in the companion 
paper \cite{discreteness2_mjbm},
that the same non-commutativity of the limits is
observed in the evolved systems. This in fact is
just a specific example of a well-known fact about
the validity of continuum Vlasov dynamics to describe
a system with long range interactions 
\cite{spohn, braun+hepp}. 
In this context it is known and well documented in 
certain systems that the continuum limit is approached as 
$N \rightarrow \infty$ keeping the time of evolution
fixed, while taking the time to infinity first one
diverges from the collisionless limit
(see, e.g., \cite{yamaguchi_etal_04}). Lowering the 
initial amplitude of a NBS increases the time of 
evolution (up to a given time), and thus the behavior
we are inferring from the analysis of the IC corresponds
to this same one. 

These comments on the continuum limit are also of
practical relevance, as they tell us how one should
study convergence to this limit numerically (in order
to understand the precision of results). It follows from
what we have just discussed that it is best to keep
$z_i$ fixed as the particle density is increased.
Further the continuum limit can only be defined clearly
in the presence of a cut-off in the input PS, with
the continuum limit being approached when the interparticle
distance is decreased well below the inverse of this scale. 
In most of the numerical studies in the literature on discreteness
in cosmological NBS these points have not been taken into 
account~\footnote{An exception is some of the cited work of Melott et al..
Some sets of simulations are compared in which only the particle
density is varied, keeping both the initial amplitude and 
the cut-off in the input PS fixed in units of the box size.}.
Indeed we note that the very widely used standard software
package COSMICS for generating IC \cite{bertschingercode} fixes 
automatically the initial red-shift of the simulation when
the physical particle density is given, rather than leaving
it as a free parameter, making such controlled tests difficult.
Indeed if no cut-off is imposed in the input PS, the criterion
used to fix the red-shift makes it increase 
with the particle density. These points will be further
discussed in forthcoming work.

We recall finally that our results on the limitations of the 
use of the algorithm for very blue spectra are of relevance
to some studies in the literature of gravitational evolution
from such spectra. Specifically we note their usefulness in
understanding quantitatively results in \cite{melott+shandarin_k4}
and \cite{bagla+padmanabhan_97}. These studies consider 
gravitational N body simulations (in two and three dimensions,
respectively) starting from IC generated on a lattice using
the standard algorithm discussed here, taking input theoretical 
PS with vanishing initial power in some range of small $k$: in 
\cite{melott+shandarin_k4} a top-hat PS is used, while in
\cite{bagla+padmanabhan_97} a gaussian centred on a chosen
wavenumber. In both cases our results show that there is 
a term proportional to $k^4$ induced at small $k$ {\it already 
in the IC}, which will dominate at small $k$. The explicit expression 
for this term, which arises at second order in the expansion of the 
continuum piece $P_c(k)$ of the full PS, is given 
in Eq.~(\ref{pc-second-four}). In \cite{melott+shandarin_k4} 
the dominant contribution from the $k^4$ term in the IC at small 
$k$ is observed numerically,  and indeed the authors relate it 
(as discussed in Sect.~\ref{Glass pre-initial conditions} above) 
to Zeldovich's argument  about ``minimal power''. 
In \cite{bagla+padmanabhan_97}, on the other hand, the $k^4$ term 
is seen (and observed, as expected, to grow with an amplification 
proportional to the square of the linear growth factor) only after 
some time. The authors describe in this case the $k^4$ tail
as ``generated'' by the dynamical evolution, which is evidently 
not quite accurate as the term is in fact present (albeit at
lower amplitude) already in the IC.

    
We are indebted to Andrea Gabrielli for extensive discussions
and explanations of his results reported in \cite{andrea}. 
We thank Thierry Baertschiger and Francesco Sylos Labini for 
numerous useful conversations, and Alvaro Dominguez and Alexander 
Knebe for helpful comments on the first version of this paper.
We are indebted to an anonymous referee for pointing out an
important error in the first version of this paper.  

\appendix

\section{Properties of the  expansion of $P_c(\mathbf{k})$}
\label{expansion-Pc}
In this appendix we study in more detail the perturbative
expansion used in the paper of $P_c(\mathbf k)$, Eq.~\eqref{P_cont}.

To simplify our analysis we will take the 
function $d_{ij}(\mathbf r)$ [defined in Eq.~\eqref{definition-d}] 
to be diagonal and isotropic, i.e.,  $d_{ij}(\mathbf r)=d(r)\delta_{ij}$.
This allows us to obtain simple analytical results, which are
exact in one dimension and which we expect to be valid only with 
minor modifications in three dimensions.

Expanding Eq.~\eqref{P_cont} in powers of $d(r)$ we have
\be
\label{pc-series-gen}
P_c(k)=\sum_{m=1}^\infty(-k^2)^{m}\int_{\mathbb R^d} d^d r e^{-i \mathbf{k}\cdot\mathbf{r}}[d(r)]^m.
\ee
We will suppose a theoretical PS in the form of
Eq.~\eqref{PS-generalIC} (and assume that
$g_{ij}(\mathbf{k})=\delta_{ij} g(k)=\delta_{ij} P_{th}(k)/k^2$)  

\subsection{Case $-d<n<-d+2$}
We can work in this case without the UV cut-off in the PS,
since $d(r)$ is well defined without it 
[cf. Eqs.~\ref{definition-d}].  Their evaluation gives
\bea
d(r) &=& -\frac{A}{\pi}\Ga(n-1)\sin\left(\frac{n\pi}{2}\right) r^{1-n}
\qquad [d=1] \\
     &=& \frac{1}{\pi^2}\Ga(n)\sin\left(\frac{3n\pi}{2}\right)r^{-1-n}
\qquad [d=3] 
\eea
The integrals in \eqref{pc-series-gen} are then divergent as 
$r \rightarrow \infty$, but defined in the sense of distributions. 
Evaluating them we obtain
\be
\label{pc-series}
P_c(k)=\sum_{m=1}^\infty P_c^{(m)}(k)=
\sum_{m=1}^\infty A^m a_m  k^{m(n+d)-d}\,.
\ee
where
\bea
\label{am1D}
\nonumber
a_m&=&-A\frac{2 \pi^{-m}}{m!} \sin\left(\frac{1}{2}m\pi(n-1)\right)\Ga(1+m-mn)\\
&\times&\left(\Ga(n-1)\sin\left(\frac{n\pi}{2}\right)\right)^m \qquad
[d=1]  \\
\label{am3D}
&=&A\frac{2^{2-m}\pi^{1-2m}}{m!} \Ga(2-m(1+n)) \\\nonumber&\times& \sin\left(\frac{1}{2}m(n+1)\pi\right)
\left(\Ga(n)\sin\left(\frac{n\pi}{2}\right)\right)^m
\qquad [d=3] 
\eea

\begin{figure}
\begin{center}
\resizebox{7cm}{!}{\includegraphics*{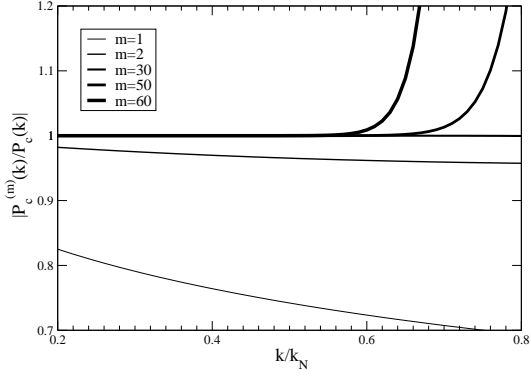}}
\caption{Ratio of the series \eqref{pc-series} summed up to
the $m$-th term to the exact (numerically evaluated) $P_c(k)$. 
We observe that there is first clear convergence (i.e. approach to unity)
as $m$ increases, and then divergence at larger values of $m$.
\label{convergence}}
\end{center}
\end{figure}

We note that the expansion \eqref{pc-series} is in fact 
an {\em asymptotic expansion}, i.e., it is strictly
divergent, but if an appropriate finite number of terms 
are taken, for any given $k$, it approximates closely
the well defined function Eq.~\eqref{P_cont}.
This behavior is shown in Fig.~\ref{convergence}, in which is 
plotted the ratio of the series \eqref{pc-series} summed
up to the $m$-th term, and $P_c(k)$. We see that for
the ratio first converges to unity as $m$ increases,
but then diverges at progressively smaller k for $m>30$.

$P_{th}(k)$ is well approximated by $P_c(k)$ if
\be
A k^n\gg A^2 a_2 k^{2(n+d)-d},
\ee
i.e., for
\be
\label{cond-k}
A a_2 k^{n+d}\ll1.
\ee
It can be checked using Eqs.~\eqref{am1D} or \eqref{am3D} that
$a_m/a_{m-1}$ is of order unity for small $m$, so that 
Eq.~(\ref{cond-k}) corresponds to the criterion 
Eq.~(\ref{validity-criterion-2}) given in the paper.
Note that we can rewrite Eq.~\eqref{cond-k} in terms of the 
variance of mass in spheres of the theoretical fluctuations, 
using the approximation (e.g. \cite{glasslike,book}):
\be
\si^2(R)=b k^d P_{th}(k)|_{k=R^{-1}},
\ee
where the coefficient $b$ is of order unity. The condition for
faithful representation of the PS of the input model at wavenumber
$k$ can thus be written:
\be
\label{cond-var-case1}
 \si^2(R)|_{k=R^{-1}}\ll1.
\ee

\subsection{The case $-d+2<n<\infty$}
In this case we must include the UV cut-off in the PS, in order
that $d(r)$ be well defined. The latter is then not a simple
power-law at all scales as in the precedent case, and 
we are unable to compute analytically the terms of the series 
\eqref{pc-series-gen}. We can, however, compute very
simply the first corrections to $P_{th}(k)$. Since $g(0)$
is finite (for $-d+2<n<\infty$), we can rewrite 
Eq.~\eqref{pc-series-gen} as 
\be
\label{pc-fact}
P_c(k)=e^{-k^2 g(0)}\int_{\mathbb R^d} d^d r e^{-i\mathbf{k}\cdot\mathbf{r}}\left( e^{k^2 g(r)}-1\right),
\ee
where we have used the identity \cite{andrea}:
\be
(2\pi)^d\delta(\mathbf{k})=(2\pi)^d e^{-k^2 g(0)}\delta(\mathbf{k})=e^{-k^2 g(0)}\int d^dr e^{-i\mathbf{k}\cdot\mathbf{r}}.
\ee
Expanding first the exponential containing $g(r)$ in Eq.~(\ref{pc-fact}) we 
obtain
\bea
\label{pc-second}
&&P_c(k)= e^{-k^2 g(0)}\\\nonumber
&&\times\left(k^2\tilde g(k)+k^4\int_{\mathbb R^d} d^d r [g(r)]^2 e^{-i\mathbf{k}\cdot\mathbf{r}}+\mathcal{O}(k^6[g(r)]^3)\right).
 \eea 
Expansion of the exponential pre-factor then gives
\bea
\label{pc-second-four}
&&P_c(k)\simeq P_{th}(k)+\\\nonumber
&&k^4\left(\int_{\mathbb R^d} d^d r [g(r)]^2 e^{-i\mathbf{k}\cdot\mathbf{r}}-g(0)\tilde g(k)\right).
\eea
By dimensional analysis one can see that the integral
in Eq.~\eqref{pc-second-four} scales as $(c_1+c_2k^{d+2n-4})$, where
$c_1$ and $c_2$ are non-zero constants. For the range of index $n$ 
considered it follows that:
\begin{itemize}
\item For $-d+2<n<2$, the dominant correction to $P_{th}$ comes from 
the term $\propto g(0)\tilde g(k)$ and therefore:
\be
\label{condi-second}
P_c(k)\simeq k^2\tilde g(k)-k^4 g(0)\tilde g(k).
\ee
It follows that the condition for a faithful representation
of the theoretical PS ($P_{th}(k)=k^2 g(k)$) is
\be
\label{cond-case2}
g(0)k^2\ll1,
\ee
which corresponds to the condition Eq.~(\ref{validity-criterion-2}). 
For a sharply cut-off theoretical PS
\be
\label{PS-theo}
P_{th}(k) = \left\{ \begin{array}{ll}
         Ak^n & \mbox{for $k\le k_c$};\\
         0 & \mbox{otherwise.}\end{array} \right. 
\ee
one has 
\bea
g(0) &=& \frac{A k_c^{n-1}}{\pi(n-1)} \qquad [d=1] \\
     &=& \frac{A k_c^{n+1}}{2 \pi^2(n+1)} \qquad [d=3]
\eea
Dropping the numerical factors, for simplicity, the condition
Eq.~(\ref{cond-case2}) can be written as
\bea
\Delta_{\rm th}^2(k) \ll && \left(\frac{k}{k_c}\right)^{n-1} \qquad [d=1] \\
     && \left(\frac{k}{k_c}\right)^{n+1} \qquad [d=3] \\
\eea
Since we are considering the case $n>-d+2$ here, this means that
Eq.~(\ref{cond-case2}) is, for $k<k_c$, a more restrictive criterion
than that found in the previous case. However, since $\Delta_{\rm th}^2(k)$
is a monotonically increasing function of $k$ up to $k_c$, the 
two conditions are essentially equivalent in cosmological
NBS, in which one generically imposes a cut-off around $k_N$.
We note further that the condition is then also equivalent to 
\be
\label{cond-var-case2}
 \si^2(R)|_{k_N=R^{-1}}\ll1,
\ee
which is equivalent to \eqref{cond-var-case1}
(since $\si^2(R)$ is in this case also a monotonically
decreasing function of $R$).

\item For $2<n<4$ the main correction comes from the integral in Eq.~\eqref{pc-second-four}:
\be
\label{condi-second-n>2}
P_c(k)\simeq k^2\tilde g(k)+\frac{1}{2}k^4\int d^dr [g(r)]^2 e^{-i\mathbf{k}\cdot\mathbf{r}}.
\ee
For the sharply cut-off theoretical PS of Eq.~(\ref{PS-theo}),
the integral can be evaluated analytically in the limit 
$\mathbf{k} \to 0$. This gives
\bea
P_c(k) &\simeq& Ak^n+A^2\frac{k^4}{2\pi}\frac{k_c^{2n-3}}{2n-3}
\qquad [d=1] \\
      &\simeq& Ak^n+A^2\frac{k^4}{2\pi^2}\frac{k_c^{2n-1}}{2n-1}
\qquad [d=3] \,.
\eea
Up to numerical factors of order unity the leading correction
is the same as in the previous case, and thus the same criteria
apply for the validity of the perturbative expansion as in
the previous case.

\item For $n>4$ the resulting PS is dominated by the $k^4$ correction.
The full expression for $P_{c}(k)$ therefore does not approximate
$P_{th}(k)$ at sufficiently small $k$.

\end{itemize}

\section{Discreteness corrections to the PS}
\label{Discreteness corrections to the PS}

We analyse further in this appendix the full expansion to
all orders of the exact discreteness correction in the PS
Eq.~\eqref{P_disc}. Then for the specific case of an input
PS with $-d<n<-d+2$, and no UV cut-off, we can evaluate
the expression analytically in one dimension. This
gives, in particular, an  analytic expression for the 
coefficient of the leading contribution, proportional to
$k^2$ and allows a precise determination of the range of
$k$ in which this term is sub-dominant with respect to
the input PS.

Expanding the exponential in Eq.~\eqref{P_disc} we have 
\bea
\label{pd-series-gen}
\Delta P_d(k) &\equiv& P_d(k) - P_{in}(k) \\\nonumber
&=& \sum_{m=1}^\infty(-k^2)^{m}\int_{\mathbb R^d} d^d r
e^{-i\mathbf{k}\cdot\mathbf{r}}[d(x)]^m \xi_{in}(\br) \,.
\eea
which can be rewritten as 
\be
\label{pd-dev}
\Delta P_d(k) =\frac{1}{(2\pi)^d} \sum_{m=1}^\infty(-k)^{2m} \int_{\mathbb R^d} d^d q D^{(m)}(\bq)P_{in}(\mathbf{q}+\mathbf{k})
\ee
where 
\be
D^{(m)}(\mathbf{k}):= {\rm FT}[(d(x)^m],
\ee
where ${\rm FT}$ denotes the FT as defined in Eq.~\eqref{FT-def}. 
For a pre-initial simple cubic lattice this gives
\be
\label{pd-dev-lat}
\Delta P_d(k)=\sum_{m=1}^\infty (-k)^{2m}\sum_{\mathbf{q}\ne 0}  D^{(m)}(\mathbf{q}+\mathbf{k})
\ee
where
\be
\mathbf{q}=k_N \mathbf{n},
\ee
and  $\mathbf{n}$ are triple integers. The 
smallest $\mathbf{q}$ in the sum \eqref{pd-dev-lat} is the Nyquist 
frequency, so that the leading term at small $k$ is 
the one we discussed in 
Sect.~\ref{subsection-The leading non-trivial discreteness correction},
proportional to $k^2$.

In one dimension all the terms in the series \eqref{pd-dev-lat}
may be calculated analytically, for the case $-d<n<-d+2$ without
a UV cut-off. The leading $k^2$ term 
is
\be
P_d(k)=2A k_N^{n-2} \zeta(2-n) k^2+\mathcal{O}(k^3).
\ee
We can then estimate the scale $k$ up to which this
term is sub-dominant compared to the input spectrum:
\be
k\lsim \left(2 \zeta(2-n)\right)^{1/n-2} k_N .
\ee
We note that this scale is independent of the amplitude of the PS
(and for $n=-1/2$, $k\lsim 4.2 k_N$).  This result is completely
in line with the numerical calculations of the contributions of these 
terms for various other cases presented in  Sect.~\eqref{subsection-The leading non-trivial discreteness correction}.

\section{Analytical results in one dimension}
\label{1d}
In this appendix we give some simplified analytic expressions
for the exact PS, mass variance and two point correlation function 
in one dimension. We have made use of these expressions in our numerical
study of various input PS in Sect.~\ref{comparison}.

We recall first the correlation properties of a
simple cubic lattice (in $d$ dimensions for generality) 
which we will take as the ``pre-initial'' distribution
in what follows. For the reduced two point correlation
function one has 
\be
\label{xi_lattice}
\ti\xi_{lat}(\mathbf r_1,\mathbf r_2)=\lan\overline{\rho(\mathbf r)\rho(\mathbf r')}\ran-1=\sum_{\mathbf l}\delta\left(\mathbf r_1-\mathbf r_2-\mathbf l\right)-1,
\ee
where $\mathbf l$ is a generic displacement vector of the lattice.
The expression Eq. (\ref{PS_lattice1}) is simply the Fourier
transform of this expression. 

Let us now consider the case of one dimension. 
To compute the variance we use its expression 
as a function of the PS (see \cite{glasslike}):

\be
\label{variance_PS}
\sigma^2(R)=\frac{1}{2\pi}\int_{-\infty}^{+\infty}dk\,\left(\frac{\sin(kR)}{kR}\right)^2
P(k) \ee
or, equivalently, as a function of the correlation function:
\bea
\label{variance_xi}
&&\sigma^2(R)=\frac{1}{8R^2}\int_{-\infty}^{+\infty}dx\,\tilde\xi(x)\times\\\nonumber
&\times&\left[-2x\theta(x)+(x-2R)\theta(x-2R)+(x+2R)\theta(x+2R)\right],
\eea
where $\theta(x)$ is the Heaviside function. Using Eqs.
(\ref{variance_PS}) or (\ref{variance_xi}) with (\ref{PS_lattice1}) or
(\ref{xi_lattice}) respectively, we obtain the following result 
for the variance of a lattice with grid spacing equal to unity :
\be
\si^2_{lat}(R)=\sum_{m=-\infty,\ne0}^{+\infty}\left(\frac{\sin(2\pi mR)}{2\pi mR}\right)^2.
\ee
As anticipated in the previous section 
we obtain the same limiting behavior of the
variance at large scales as  for a homogeneous 
and isotropic distribution with PS $P(k)\sim k^n$ and 
$n>1$, i.e.,  $\si^2(R)\sim 1/R^{d+1}$ with $d=1$.

We now compute an expression for the PS directly
from (\ref{P_3d}), for the case of a
one-dimensional system and a ``pre-initial'' lattice 
configuration. Using
Eq. (\ref{xi_lattice}) and rearranging terms we obtain:

\begin{eqnarray}
\label{PS_lattice}
P(k)&=& \exp(-k^2 g(0)) \sum_{-\infty,l\ne0}^{+\infty}\delta(k-2\pi
l)\\\nonumber 
&+&\sum_{l=-\infty}^{+\infty} e^{-ikl}[\exp(-k^2
d(l)))-\exp(-k^2 g(0))],
\end{eqnarray}
where $d(x)\equiv g(0)-g(x)$. The first term on the right hand side
of Eq. (\ref{PS_lattice}) contains all the divergent terms in the PS. The
second term is a regular function of $k$ which has the behavior
$P(k)\sim k^2 g(k)$ at small $k$ if $g(k)\sim k^\alpha$ with
$\alpha<0$ and $ P(k)\sim k^2$ if $\alpha>0$, unless 
$\sum_{l=-\infty}^{+\infty} g(l)=0$, in which case $P(k)\sim k^2 g(k)$
also for $\alpha>0$.

Performing a Fourier transform of Eq. (\ref{P_3d}) we obtain
the correlation function in the form
\begin{eqnarray}
\label{xi_andrea}
\nonumber \tilde{\xi}(x)&=&\frac{1}{2\pi}\int_{-\infty}^{+\infty}
dx'\,\sqrt\frac{\pi}{d(x')}e^{-(x-x')^2/4 d(x')}\times\\
&\times&\left(1+\tilde{\xi}_{in}(x')\right)-1.
\end{eqnarray}
Note that in the limit that no displacements are applied (i.e. $d(x)\to0$), 
the argument of the integral is $\delta(x-x')$.  Thus we recover
explicitly for small displacements
$\tilde\xi(x)\simeq\tilde\xi_{in}(x)+\dots$. Substituting  
Eq. (\ref{xi_lattice}) in Eq. (\ref{xi_andrea}) we then obtain
the result for the specific case of a ``pre-initial'' lattice 
configuration:

\begin{equation}
\tilde{\xi}(x)=-1+\sum_{l=-\infty}^{+\infty}\sqrt{\frac{1}{4\pi
d(l)}}e^{-(x-l)^2/4d(l)}.
\end{equation}
To obtain the variance we use the same procedure. Using, for example,
Eq. (\ref{variance_PS}) with Eq. (\ref{P_3d}) we get:

\begin{eqnarray}
\nonumber \sigma^2(R) & = & -1+\frac{1}{4\sqrt{\pi}
R^2}\int_{-\infty}^{+\infty}dx\,(1+\tilde{\xi}_{in}(x))\sqrt{\,d(x)}\times\\\nonumber
&\times&\left[h(x,2R)+h(x,-2R)-2h(x,0)\right]\\ & + &
\frac{1}{8R^2}\int_{-\infty}^{+\infty}dx\,(1+\tilde{\xi}_{in}(x))\times\\\nonumber
&\times&\left[-2f(x,x)+f(x-2R,x)+f(x+2R,x)\right]
\end{eqnarray}
where
\begin{equation}
f(x,y)=x\,\mathrm{erf}\left(\frac{x}{2\sqrt{d(y)}}\right),\,\,
h(x,y)=e^{\frac{-(x+y)^2}{4 d(x)}}.
\end{equation}
Expanding at small $d(x)$ it is possible to obtain 
also explicitly an expression of the form $\si^2(R)=\si_{lat}^2(R)+...$. In the specific case of an initial lattice distribution the variance can be written:

\begin{eqnarray}
\label{sigma_lat}
\nonumber
&&\sigma^2(R)  = 
-1+\frac{1}{4\sqrt{\pi} R^2}\sum_{l=-\infty}^{+\infty}\sqrt{\,d(l)}\times\\
&\times&\left[h(l,2R)+h(l,-2R)-2h(l,0)\right]\\\nonumber
& + & 
\frac{1}{8R^2}\sum_{l=-\infty}^{+\infty}\left[-2f(l,l)+f(l-2R,l)+f(l+2R,l)\right].
\end{eqnarray}



\end{document}